\documentclass[12pt]{iopart}
\usepackage{iopams}  
\usepackage{afterpage}
\usepackage{graphicx}  
\usepackage{amssymb}

\begin{document}

\title[Fermionic RG methods for inhomogeneous Luttinger
liquids]{Fermionic renormalization group methods for 
transport through inhomogeneous Luttinger liquids}

\author{V Meden$^1$, S Andergassen$^2$, T Enss$^3$,
H~Schoeller$^1$, \\ and K Sch\"onhammer$^4$}
\address{$^1$ Institut f\"ur Theoretische Physik A, RWTH Aachen,
  52056 Aachen, Germany}
\ead{meden@physik.rwth-aachen.de}
 
\address{$^2$ Institut N\'eel, Centre National de la Recherche Scientifique and
   Universit\'e Joseph Fourier, BP 166, 38042 Grenoble, France} 

\address{$^3$ Dipartimento di Fisica, Universit\`a di Roma ``La Sapienza'',
  Piazzale Aldo Moro 2, I-00185 Roma, Italy}

\address{$^4$ Institut f\"ur Theoretische Physik, Universit\"at
  G\"ottingen, Friedrich-Hund-Platz 1, 37077 G\"ottingen, Germany} 

\begin{abstract}

We compare two fermionic renormalization group methods which have been
used to investigate the electronic transport properties of one-dimensional
metals with two-particle interaction (Luttinger liquids) and 
local inhomogeneities. The first one is a poor man's method setup to
resum ``leading-log'' divergences of the effective
transmission at the Fermi momentum. Generically the resulting 
equations can be solved
analytically. The second approach is based on the functional
renormalization group method and leads to a set of differential 
equations which can only for certain setups and in limiting cases 
be solved analytically, while in general it must be integrated 
numerically. Both methods are claimed to be applicable for 
inhomogeneities of arbitrary strength and to capture effects
of the two-particle interaction, such as 
interaction dependent exponents, up to leading order. 
We critically review this for the simplest case of a single impurity. 
While on first glance the poor man's approach seems to describe the
crossover from the ``perfect'' to the ``open chain fixed point'' 
we collect evidence that difficulties may arise close to the 
``perfect chain fixed point''. Due to a subtle relation between the
scaling dimensions of the two fixed points this becomes apparent only
in a detailed analysis. In the functional
renormalization group method the coupling of the different scattering
channels is kept which leads to a better description of the underlying
physics.     

\end{abstract}

\pacs{71.10.-w, 71.10.Pm, 73.21.Hb}

\maketitle

\section{Introduction}
\label{intro}

The sensitivity of one-dimensional (1D) correlated electron systems --
so called Luttinger liquids (LL) \cite{Schoenhammer05} -- to
local single-particle inhomogeneities (e.g.~a single impurity)
leading to a (back-) scattering of electrons with
momentum transfer $2 k_F$ (with $k_F$ being the Fermi momentum) was
recognized several decades ago \cite{LutherPeschel, Mattis, ApelRice,
  Giamarchi}. The density-density response function at zero frequency
shows a power-law singularity when the momentum transfer $q$
approaches $2 k_F$ with an exponent determined by the Luttinger liquid
parameter $0<K<1$ (for repulsive interactions; $K=1$ in the 
noninteracting case). It was computed for 
a specific field-theoretical model -- the Tomonaga-Luttinger (TL) 
model -- after mapping the interacting fermionic problem onto a 
free bosonic Hamiltonian using the bosonization method \cite{Herbert}. 
The TL model is known to describe the low-energy physics of a large 
class of 1D metals \cite{Schoenhammer05}.  

At the beginning of the 90's it became apparent that the
realization of sufficiently well defined 1D quantum wires and thus 
the possibility to experimentally investigate LL physics would only
be a matter of a few more years of research. This was one of the
reasons for a renewed theoretical interest in such 
systems. The understanding of the 
role of local impurities was further sharpened by renormalization 
group (RG) studies within the bosonized model \cite{KaneFisher,Furusaki}.  
In these studies the TL model was supplemented in two ways. In a first
step a term solely describing the $k_F$ to $-k_F$ (and vice versa) 
backscattering (of amplitude $V_b$) was added to the Hamiltonian. 
Using bosonization this leads to a term $ \propto 
\sin{[\phi(0)]}$, with $\phi(x)$ being the bosonic 
field and $x=0$ the  position of the
impurity. The resulting field-theoretical model is known as 
the local sine-Gordon model. In a RG procedure perturbative in 
$V_b$ it was shown that $V_b$  grows and is thus a relevant 
perturbation (for repulsive interactions with $0<K<1$). 
At a sufficiently low energy scale the
perturbative RG breaks down. In a second step the opposite case of two
semi-infinite TL chains with a weak link of amplitude $t_{\rm wl}$ 
was studied. This was shown to scale to zero. Based on these insights 
and an exact solution at $K=1/2$ it was speculated that for 
repulsive interactions even a weak single impurity will scale
towards the strong impurity limit implying that at low energies
observables show power-law scaling with exponents characterized by the
scaling dimension of the ``open chain fixed point'' $1/K$. E.g.~the
local spectral function $\rho(x\approx 0,\omega)$ close to the
impurity will vanish as $\omega^{1/K-1}$ and the conductance $G$ 
across the impurity as 
\begin{eqnarray}
\label{stronglimit}
G \sim T^{2(1/K-1)} \; , 
\end{eqnarray}
with $T$ being the temperature.
 For small $V_b$ the corrections to 
``perfect'' $G=e^2/h$ conductance \cite{footnote} 
are given by the scaling dimension $K$ 
of the (unstable) ``perfect chain fixed point'', 
\begin{eqnarray}
\label{weaklimit}
G- \frac{e^2}{h} \sim - V_b^2 \, T^{2(K-1)} \; . 
\end{eqnarray}
Within the local sine-Gordon model the above conjecture was later confirmed 
using numerical \cite{Moon} and exact \cite{Fendley} methods.    

The physics of spinless inhomogeneous LLs and systems with spin is
quite similar in many respects and for simplicity we here mainly focus
on the spinless case. 
The scaling exponents just given are slightly different if the spin is 
included.
Towards the end of this article we briefly explain
how the spin can be included in the two  fermionic RG approaches
described here and discuss a novel feature which results from it.  

In parallel to the bosonization approach 
a fermionic poor man's RG method was developed which does
not rely on the mapping onto a bosonic field theory \cite{poor}. 
In this approach the single-particle impurity scattering problem is
first solved exactly within the 1D electron gas model. Next, 
the Hartree and Fock terms [leading order perturbation theory 
in the two-particle interaction $U(x)$] 
are computed with the scattering states. The Hartree diagram 
produces a potential term which oscillates as $\cos{(2 k_F x)}$, decays as 
$1/x$, and has  a prefactor given by the reflection amplitude of the (bare) 
impurity. Scattering off the Hartree potential
leads to a logarithmically diverging
correction to the transmission amplitude $t(k)$ if $k \to k_F$
\begin{eqnarray}
\label{poordiv}
\delta t(k) \sim - \alpha\, \ln{\left| \frac{1}{d(k-k_F)}\right|}  \; ,  
\end{eqnarray}
where $d$ is a characteristic spatial scale of the interaction and 
\begin{eqnarray}
\label{alphadef}
\alpha = \frac{\tilde U(0)}{2 \pi v_F} -  
\frac{\tilde U(2 k_F)}{2 \pi v_F} 
\end{eqnarray}
is the relevant amplitude of the (Fourier transform of the) 
interaction, with $v_F$ being the Fermi velocity. The authors 
argue that one can iterate this procedure and produce 
divergent terms of higher order 
$\left[ \alpha \ln{\{ 1/ | d (k-k_F)  | \} }\right]^n$. 

To resum these terms a real space RG method is suggested. 
At scale $\Lambda = v_F/l$ the scattering potential 
$\sim \cos{(2 k_F x)}/x$ is considered only for $x \in [-l,l]$ 
with a spatial cutoff $l$. This way the logarithmic divergence 
is regularized. Then the region $[-l,l]$ with its renormalized 
reflection and transmission amplitudes is viewed as an 
effective barrier from which the Hartree potential and the 
Fock term for the region outside $[-l,l]$ can be computed. One next 
considers $l \to l + \delta l$ and iterates the above steps. 
This leads to a single differential equation for the transmission 
amplitude at momentum $k_F$. For the following analysis it is 
essential to note that the nontrivial spatial dependence of the 
scattering potential at each step is $\sim \cos{(2 k_F x)}/x$ 
and does not change during the RG procedure.  

It was argued that, in contrast to the bosonic RG method which is
perturbative in the impurity strength, this fermionic procedure,
although being applicable only for sufficiently small $\alpha$ (weak
two-particle interaction), can be used to describe the full crossover 
from the weak to the strong impurity limit. Further down we 
challenge this statement. A detailed analysis for the single impurity
case shows that the poor man's approach heavily relies on a subtle
relation between the scaling dimensions of the ``perfect'' and ``open
chain fixed points''. 
The method is commonly regarded as being very intuitive 
and has been generalized to study transport through 
two barriers (resonant tunneling) \cite{Nazarov,Polyakov}, 
through Y-junctions of three LLs \cite{Lal}, and 
through regular networks of LLs \cite{Doucot} as well as transport 
across a LL-superconductor junction \cite{Belzig}. Remarkably, 
the lack of a formal framework from which the poor man's flow
equations can be deduced led to a dispute about their correct form for
the double barrier case \cite{Nazarov,Polyakov}.  
 
In both the fermionic RG as well as the field-theoretical description 
only the scaling of the backscattering amplitude  $k_F \to -k_F$ 
(and vice versa) was considered. Undoubtedly the backscattering 
component $V_b=V_{k_F,-k_F}$ of a 
general scattering potential with matrix elements $V_{k,k'}$ is 
(i) the one which grows the fastest when the infrared 
cutoff $\Lambda$ is lowered and is (ii) the component which at 
the end of the RG flow determines the conductance across 
the impurity. But in more elaborate RG procedures all scattering 
channels $k \to k'$ will couple and also the amplitudes $V_{k,k'}$ with 
$k$, $k'$ not equal to $\pm k_F$ will grow. Even if the
matrix elements with momentum transfer different from 
$2 k_F$ are absent in the original Hamiltonian generically they will 
be generated during the RG procedure.
One might speculate that this feedback will lead to 
intermediate fixed points not captured by the local sine-Gordon 
model (as no other matrix elements than
$V_{k_F,-k_F}$ are kept in the model) and the poor man's RG (as the
feedback is neglected). Motivated by this reasoning 
an alternative fermionic RG 
procedure was developed starting from the rather general 
functional RG (fRG) framework \cite{SalmhoferHonerkamp,lecturenotes}. 
It allows for a description of the full functional form of the 
renormalized effective impurity potential. Using truncated fRG flow equations 
it was shown that indeed {\it no} intermediate fixed points are realized 
\cite{frg1,frg2,frg3}. As a further advantage this RG method 
can directly be applied to microscopic lattice models and does not
require a mapping onto the 1D electron gas or an effective field 
theory. It is rather straightforward to extend the fRG approach 
to study resonant transport through a double barrier \cite{frg3}, 
transport through a wire with three and four impurities \cite{frg3a},
and junctions of more than two LLs \cite{frg4a,frg4b}.

In the fRG formalism one derives an infinite hierarchy of coupled
differential flow equations for the cutoff dependent one-particle
irreducible vertex functions. In practical applications this 
hierarchy must be
truncated. To study the physics of local inhomogeneities in spinless 
LLs up to leading order in the two-particle
interaction it is sufficient to keep the flow of the frequency
independent self-energy (one-particle vertex) \cite{frg1}. The 
quantitative agreement to the results derived by bosonization in the
limits of weak and strong impurities as well as to numerical results 
obtained for small systems can be improved if the flow of the static
part of the effective two-particle interaction (two-particle vertex)
is kept \cite{frg2,frg3}. Details are given below. The frequency
independent self-energy can be interpreted as an effective,
renormalized impurity potential. For strong bare impurities it
spatially oscillates with frequency $2k_F$ and decays as the inverse
distance from the inhomogeneity up to a scale $v_F/T$, beyond which it
dies off exponentially. Scattering off this effective impurity then
leads to the power-law decay of the conductance as a function 
of $T$. For weak bare
impurities and not too low temperatures the oscillatory effective 
potential only decays with exponent $K$. For lower temperatures a
crossover to the strong impurity behavior can be observed and the
system flows to the ``open chain fixed point''.

Below we give more details on the truncated fRG procedure and the basic
equations for the poor man's RG (Sec.~\ref{methods}). We discuss the
most important results and compare the outcomes of the two fermionic 
RG approaches (Sec.~\ref{results}). It is shown that the truncated 
fRG can be used to describe the full crossover from the weak to the 
strong impurity limits while the poor man's RG seems to have
difficulties close to the ``perfect chain fixed point''. We show that a 
single RG equation -- as derived in the poor man's RG for the 
effective transmission at momentum $k_F$ -- is in general 
not enough to describe the crossover behavior. 
We then briefly comment on how the spin degree of freedom is included
in both methods (Sec.~\ref{spin}). Finally, the comparison of the
two fermionic RG methods for the
single impurity case is summed up and briefly extended to 
Y-junctions of three LLs (Sec.~\ref{summary}). 
In the Appendix we discuss the additional approximations necessary to
derive the poor man's RG equation within the fRG framework.

\section{The methods}
\label{methods}

In this section we present the basic equations of the fRG approach and
the poor man's RG. We focus on the simplest situation of a single
impurity in an otherwise ``perfect'' chain which is coupled ``adiabatically'' 
to two semi-infinite noninteracting leads \cite{footnote}. 

\subsection{The functional RG approach}

The fRG was recently introduced as a powerful new  
tool for studying interacting Fermi systems \cite{SalmhoferHonerkamp}. 
It provides a systematic
way of resumming competing instabilities and goes
beyond simple perturbation theory even in problems which are not 
plagued by infrared divergences.
In our application the interacting quantum
wire will be coupled to noninteracting leads as it is the case 
in systems which can be realized in experiments. The leads are modeled
as 1D tight-binding chains with hopping matrix element $t=1$.
As we are interested
in the effect of a single impurity in the bulk part of the wire and
not the role of the contacts the latter are modeled as being perfect,
that is the conductance through the interacting system of length $N$
in the absence of the impurity is $e^2/h$ (up to our numerical
accuracy). This implies that the interaction must be turned on
spatially smoothly close to the contacts 
(``adiabatic contacts'') \cite{frg3,footnote}. 
Before setting up the fRG scheme we integrate out the noninteracting
leads using a standard projection technique so that we only have 
to deal with an interacting system of size $N$ \cite{frg3}.
The energy scale $\delta=v_F/N$ related to the length of the
interacting part will provide an infrared cutoff for power-law
scaling as a function of other energy variables. 

In the fRG procedure the noninteracting propagator 
$\mathcal{G}_0$ is replaced by a propagator 
depending on an infrared cutoff $\Lambda$. Specifically,
we use 
\begin{eqnarray}
\label{cutoffproc}
{\mathcal G}_0^{\Lambda}(i \omega_n) = \chi^\Lambda(\omega_n) 
{\mathcal G}_0(i \omega_n)
\end{eqnarray}
with $\Lambda$ running from $\infty$ down to $0$. 
The cutoff function $\chi^\Lambda$ is chosen such that 
the fermionic Matsubara frequencies $|\omega_n| \lessapprox \Lambda$ are 
cut out (for details see the Appendix A
of Ref.~\cite{frg5}). Note that ${\mathcal G}_0$ contains the 
frequency dependent self-energy contribution from the leads 
\begin{eqnarray}
\label{leadpotdef}
 \Sigma_{j,j'}^{\rm lead}(i\omega_n)  = 
 \frac{i\omega_n+\mu}{2} \left( 1 - 
 \sqrt{1 - \frac{4}{(i\omega_n+\mu)^2}} \, \right)
 \delta_{j,j'} \left( \delta_{1,j} + \delta_{N,j} \right) \; ,
\end{eqnarray}
with $j,j' \in [1,N]$ indicating the lattice site 
and the filling $n$-, interaction $U$-, and $T$-dependent 
chemical potential $\mu = \mu(n,U,T)$. It is chosen such that 
the entire system (wire and leads) has the desired filling at any $U$
and $T$.  Using  ${\mathcal G}_0^\Lambda$ in the 
generating functional of the irreducible vertex 
functions and taking the derivative with respect to 
$\Lambda$ one can derive an exact, infinite hierarchy of coupled differential
equations for the vertex functions, such as the self-energy
$\Sigma^\Lambda$ and the irreducible two-particle
interaction $\Gamma^{\Lambda}$. In 
particular, the flow of the self-energy (one-particle
vertex) is determined by $\Sigma^\Lambda$ itself and 
the two-particle vertex, while the flow of 
$\Gamma^{\Lambda}$ is determined by $\Sigma^\Lambda$, $\Gamma^{\Lambda}$, and 
the flowing three-particle vertex $\Gamma_3^{\Lambda}$.
The latter is computed from a flow equation involving
the four-particle vertex, and so on.
At the end of the fRG flow $\Sigma^{\Lambda=0}$ is the self-energy $\Sigma$ 
of the original, cutoff-free problem we are interested in.
A detailed derivation of the fRG flow equations for a general quantum
many-body problem which only requires a basic knowledge of the
functional integral approach to many-particle
physics and the application of the method for a
simple problem are presented in Ref.~\cite{lecturenotes}.   

In practical applications the hierarchy of flow 
equations has to be truncated and $\Sigma^{\Lambda=0}$ only provides 
an approximation for the exact $\Sigma$. As a first approximation 
we neglect the three-particle vertex.
The contribution of $\Gamma_3^{\Lambda}$ to $\Gamma^{\Lambda}$ is small
as long as $\Gamma^{\Lambda}$ is small, because 
$\Gamma_3^{\Lambda}$ is initially (at $\Lambda=\infty$) zero 
and is generated only from terms of third order in 
$\Gamma^{\Lambda}$. Furthermore,  $\Gamma^{\Lambda}$ stays small 
for all $\Lambda$ if the bare interaction is not too large. 
This approximation leads to a closed set of equations for 
$\Gamma^{\Lambda}$  and $\Sigma^{\Lambda}$ \cite{frg3}. 

We here do not give these equations but instead implement a 
second approximation: the frequency dependent flow of the 
renormalized two-particle vertex $\Gamma^{\Lambda}$ is replaced 
by its value at vanishing external frequencies, such that 
$\Gamma^{\Lambda}$ and hence $\Sigma^{\Lambda}$ remains frequency 
independent. Since the bare interaction is frequency independent, 
neglecting the frequency dependence leads to errors only at 
second order for the self-energy, and at third order for the 
two-particle vertex at zero frequency.
For the approximate flow equations we then obtain
\begin{eqnarray}
 \frac{d}{d\Lambda} \Sigma^{\Lambda}_{1',1} =
 - \frac{1}{2\pi} \sum_{|\omega_n| \approx \Lambda} \sum_{2,2'} \,
 e^{i\omega_n 0^+} \, {\mathcal G}^{\Lambda}_{2,2'}(i\omega_n) \,
 \Gamma^{\Lambda}_{1',2';1,2} 
\label{finalflowsigma}
\end{eqnarray}
and 
\begin{eqnarray}
&& \frac{d}{d\Lambda} \Gamma^{\Lambda}_{1',2';1,2}   = 
\frac{1}{2\pi} \, 
 \sum_{|\omega_n| \approx \Lambda} \, \sum_{3,3',4,4'} 
 \Big\{ \frac{1}{2} \,  {\mathcal G}^{\Lambda}_{3,3'}(i\omega_n) \, 
  {\mathcal G}^{\Lambda}_{4,4'}(-i\omega_n) 
 \Gamma^{\Lambda}_{1',2';3,4} \, \Gamma^{\Lambda}_{3',4';1,2} 
\nonumber \\ && +  {\mathcal G}^{\Lambda}_{3,3'}(i\omega_n) \,
 {\mathcal G}^{\Lambda}_{4,4'}(i\omega_n)  
 \left[ - \Gamma^{\Lambda}_{1',4';1,3} \, \Gamma^{\Lambda}_{3',2';4,2} 
        + \Gamma^{\Lambda}_{2',4';1,3} \, \Gamma^{\Lambda}_{3',1';4,2}
 \right] \Big\} \; ,
\label{finalflowgamma}
\end{eqnarray}
where the lower indexes $1$, $2$, etc.\  stand for the 
single-particle quantum numbers of the chosen basis (e.g.~the Wannier
basis or the momentum state basis) and
\begin{eqnarray}
\label{Glambdadef}
   {\mathcal G}^{\Lambda}(i\omega_n) = 
 \left[  {\mathcal G}_0^{-1}(i\omega_n) - \Sigma^{\Lambda} \right]^{-1} \; .
\end{eqnarray}
The abbreviation $|\omega_n| \approx \Lambda$ stands for taking the positive 
as well as negative Matsubara frequency with absolute value closest 
to $\Lambda$ at fixed $T$. 
At the initial cutoff $\Lambda=\infty$ the flowing two-particle vertex
$\Gamma^{\Lambda}_{1',2';1,2}$  is given by the antisymmetrized
interaction and the self-energy by the single-particle terms of the
Hamiltonian not included in ${\mathcal G}_0$, e.g.~the impurity. 

Next we specify our model for the interacting LL wire with
noninteracting leads. We use the
1D tight-binding model with nearest-neighbor hopping $t=1$ and
nearest-neighbor interaction $U_{j,j+1}$. It is given by 
 \begin{eqnarray}
\label{hdef}
\mbox{} \hspace{-1.5cm} H & = & H_{\rm kin} +  H_{\rm int}  \nonumber 
\\
\mbox{} \hspace{-1.5cm} & = & - \sum_{j=-\infty}^{\infty} \big( \,
 c^{\dag}_{j+1} c_j^{\phantom\dag} + c^{\dag}_j \, c_{j+1}^{\phantom\dag}
 \, \big) +  \sum_{j=1}^{N-1} U_{j,j+1}  \left[ n_j - \nu(n,U) 
\right]  \left[ n_{j+1} -\nu(n,U)  \right] \; ,
\end{eqnarray}
where we used standard second-quantized notation with $c^{\dag}_j$ 
and $c_j^{\phantom\dag}$ being creation and annihilation operators on site $j$,
respectively and the local density operator $n_j = c^{\dag}_j \,
c_j^{\phantom\dag}$. The filling $n$ and interaction dependent real number
$\nu$ must be determined iteratively such that at fixed common 
$\mu$ (for the leads and the wire) the interacting wire has the 
desired filling. For the mostly studied case of half-filling
particle-hole symmetry implies $\nu=1/2$.
As mentioned above the interaction $U_{j,j+1}$ is
turned on and off smoothly over a small fraction of the $N$ interacting 
lattice sites. The constant bulk value is denoted by $U$. 

The homogeneous model with a constant
interaction $U$ across all bonds 
can be solved exactly by the Bethe ansatz \cite{Haldane}. 
It shows LL behavior for all particle 
densities $n$ and any interaction strength except at half filling 
for $|U| > 2$. The $U$- and $n$-dependent 
LL parameter $K$ can be determined solving coupled
integral equations \cite{Haldane}, which in the half-filled case can be done
analytically with the result 
\begin{equation}
\label{BetheAnsatz}
 K = \left[\frac{2}{\pi} \, 
 \arccos \left(-\frac{U}{2} \right) \right]^{-1} = 1 - \frac{U}{\pi} +
{\mathcal O}(U^2) \; , 
\end{equation}
for $|U| \leq 2$. As $K$ depends on the filling it is essential to tune 
the interacting wire to the desired filling by choosing the appropriate
$\nu$ [see Eq.~(\ref{hdef})]. Only then a comparison of the
fRG results to the ones obtained by bosonization is possible.

The Hamiltonian (\ref{hdef}) is supplemented by an impurity part 
\begin{equation}
\label{impdef}
 H_{\rm imp} = \sum_{j,j'} 
 V_{j,j'}^{\phantom\dag} \; c^{\dag}_{j} \, c_{j'}^{\phantom\dag} \; ,
\end{equation}
where $V_{j,j'}$ is a static potential.
Local site impurities of amplitude $V$ at $j_0$ are given by a potential 
\begin{equation}
 V_{j,j'} = V  \, \delta_{j,j'} \, \delta_{j,j_0} 
\end{equation}
and local hopping impurities with hopping $t'$ across the link 
$j_0,j_0+1$ by
\begin{equation}
 V_{j,j'} = V_{j',j} = - (t'-1) \, \delta_{j',j+1} \, \delta_{j,j_0}
 \; .
\end{equation}

In our RG procedure we parameterize the 
static real space vertex by a flowing nearest-neighbor interaction
 \begin{equation}
\label{parametri}
 \Gamma^{\Lambda}_{j'_1,j'_2;j_1,j_2} =
  U^{\Lambda} \, (\delta_{j_1,j_2-1} + \delta_{j_1,j_2+1}) \, 
 ( \delta_{j_1,j'_1} \delta_{j_2,j'_2} - 
   \delta_{j_1,j'_2} \delta_{j_2,j'_1} ) \; .
\end{equation}
We furthermore neglect the feedback of the self-energy on the flow of
$U^{\Lambda}$. The differential equation for the renormalized bulk
interaction $U^\Lambda$ is then determined
after transforming Eq.~(\ref{parametri}) to momentum space and 
fixing the momenta on the Fermi surface \cite{frg6}. 
It reads (in the limit $N \to \infty$)
\begin{eqnarray} 
\label{flowU}
\frac{d}{d\Lambda} U^\Lambda = h(\tilde\omega_n) \, (U^\Lambda)^2
\end{eqnarray}
with 
\begin{eqnarray}
\mbox{} \hspace{-1.0cm} h(x) & = & - \frac{1}{2\pi} -
 \mbox{Re} \, \bigg[ \frac{i}{2} \, (\mu + i x) 
 \, \sqrt{1 - \frac{4}{(\mu + ix)^2}} \nonumber \\ \mbox{}
 \hspace{-1.0cm} && \times \frac
 {3i\mu^4 - 10\mu^3 x - 12i\mu^2(x^2 + 1) + 6 x^3\mu
  + 18 x \mu + 6i x^2 + i x^4}
 {\pi (2\mu + ix)(4 - \mu^2 + x^2 - 2i x \mu)^2} \,
  \bigg] , 
\end{eqnarray}
where $\tilde\omega_n$ stands for the fermionic Matsubara frequency
closest to $\Lambda$. The initial condition is $\lim_{\Lambda_0 \to
  \infty} U^{\Lambda_0} =U$.
The bond dependent renormalized interaction
$U^\Lambda_{j,j\pm1}$ 
follows from the bulk $U^\Lambda$ by multiplying with 
$U_{j,j\pm1}/U$. 
After performing
these additional approximations the flow equations for the Wannier
basis matrix elements of the self-energy read
\begin{eqnarray}
 \frac{d}{d\Lambda} \, \Sigma^{\Lambda}_{j,j} \; =&
 - & \frac{1}{2\pi} \sum_{|\omega_n|  \approx \Lambda} 
\sum_{r = \pm 1} \, U_{j,j+1}^\Lambda
  \, {\mathcal G}^{\Lambda}_{j+r,j+r}(i\omega_n) \; ,\nonumber \\
 \frac{d}{d\Lambda} \, \Sigma^{\Lambda}_{j,j \pm 1} \; =&
   & \frac{1}{2\pi} \sum_{|\omega_n|  \approx \Lambda}\,
   U_{j,j\pm1}^\Lambda \, 
  {\mathcal G}^{\Lambda}_{j,j \pm 1}(i\omega_n) \; .
\end{eqnarray} 
All other components vanish.
The self-energy enters also the right-hand side of these equations via
the full propagator ${\mathcal G}^{\Lambda}$. The initial conditions
at $\Lambda = \Lambda_0 \to \infty$ are 
\begin{eqnarray}
\Sigma^{\Lambda_0}_{j,j} =  
\left[1/2- \nu(n,U)\right] \left( U_{j-1,j} +
  U_{j,j+1} \right)  \; , \;\;\;\; \Sigma^{\Lambda_0}_{j,j\pm 1}   =  0 \; .
\end{eqnarray}
 At the end of the fRG procedure the diagonal 
part of the approximate 
self-energy $\Sigma_{j,j}$ can be interpreted as an effective impurity
potential, while $-1+\Sigma_{j,j\pm1}$ is the renormalized
hopping. 

For general parameters the set of coupled differential equations 
can only be solved numerically. This can be done for fairly 
long chains (up to $10^7$
lattice sites) as computing the right-hand side of the flow equations 
only requires the tridiagonal part of the inverse of a tridiagonal 
$N \times N$ matrix which can be obtained in order $N$ time \cite{frg6}. 
In the limiting cases of weak and strong 
impurities and neglecting the flow of the two-particle interaction
altogether the flow equations can be solved analytically at $T=0$.
This is discussed in Sec.~\ref{results}. 

The truncated fRG procedure provides an approximation
for the self-energy and thus via the Dyson equation an approximation
for the interacting one-particle Green function $\mathcal G$. 
If the inelastic (frequency dependent) processes the two-particle 
interaction generates are neglected, the conductance can be computed 
from the $(1,N)$ matrix element of the retarded Green function $\mathcal G^r$
by a generalized Landauer-B\"uttiker formula \cite{frg3} 
 \begin{equation} 
\label{LB}
G(T,\delta)=\frac{e^2}{h}\int_{-B/2}^{B/2}
\left(-\frac{df}{d \varepsilon}\right)|t(\varepsilon,T,\delta)|^2 d\varepsilon,
\end{equation}
where $f(\varepsilon)=1/(e^{(\varepsilon-\mu)/T }+1)$
is the Fermi function, $B$ the band width ($B=4$ for $t=1$), and 
 \begin{equation} 
\label{transmission}
|t(\varepsilon_k,T,\delta)|^2=4 \sin^2(k) \; 
\left|{\mathcal G}^r_{1,N}(\varepsilon_k)\right|^2 
\end{equation}    
the effective transmission. 
Because of the frequency independence of $\Sigma$ the analytic 
continuation of $\mathcal G$ to the real axis, necessary to obtain 
${\mathcal G}^r(\varepsilon)$, is straightforward. 
During the RG procedure the self-energy
and thus the propagator acquire additional dependencies on the
temperature and on the energy scale $\delta \sim 1/N$. 

Inelastic processes vanish in the limit $T \to 0$. For $T>0$ they are
absent in our approach as we assumed the self-energy to 
be frequency independent and using Eq.~(\ref{LB}) is thus consistent 
with our earlier approximations -- it presents no further approximation. 
In general inelastic processes lead to corrections to Eq.~(\ref{LB}) of
order $U^2$ and higher. It is important to note that inelastic
processes due to the two-particle interaction are also neglected in
the poor man's RG as well as in the bosonization approach. In the
latter they are disregarded because the bulk part of the local
sine-Gordon model corresponds to the TL model in which all 
infrared RG irrelevant terms such as umklapp scattering are left out. At 
finite scales such terms imply inelastic scattering.

\subsection{Solution of the poor man's RG equation}

The basic idea of the poor man's RG was already presented in the
introduction. It is set up for the 1D electron gas model. 
The method leads to the flow equation \cite{poor}
\begin{eqnarray}
\label{poormansRGequation}
\frac{d t^\Lambda}{d \ln{(\Lambda/[v_F/d])}} = \alpha
t^\Lambda (1- \left| t^\Lambda \right|^2)
\end{eqnarray}
for the transmission $t^\Lambda$ at $ k_F$, with
$d$ and $\alpha$ introduced in Sec.~\ref{intro}. 
It only depends on a single energy scale, the infrared energy cutoff
$\Lambda$, which must be contrasted to the fRG approach in which the
scales $\Lambda$, $T$, and $\delta$ appear. 
The integration of Eq.~(\ref{poormansRGequation}) 
starts at the ultraviolet cutoff $\lambda_0= v_F/d$ with the initial 
condition $t^{\lambda_0}=t_0$, where $t_0$ is the bare
transmission. To obtain the  energy or temperature dependence of the
renormalized transmission the integration is stopped at
$\Lambda=\varepsilon$ or $T$, respectively. The assumed dependence on
only one energy scale implies that the variable one is
interested in must be much larger than all the other scales. In the
fRG in contrast the interplay of the scales $\Lambda$, $T$, and
$\delta$ can be studied. The solution of Eq.~(\ref{poormansRGequation}) is 
\begin{eqnarray}
\label{poormansRGsol}
\left|t^{s}\right|^2= 
\frac{|t_0|^2 |s/\lambda_0|^{2 \alpha}}{|r_0|^{2 } +
|t_0|^2 |s/\lambda_0|^{2 \alpha} }
\end{eqnarray}
where $s$ must be substituted by the energy variable of interest
(e.g.~$s \to T$). The bare reflection probability is denoted by
$|r_0|^2=1-|t_0|^2$. The expression for the conductance follows after
multiplying the transmission (probability) by $e^2/h$.

For the upcoming analysis it is important to note that the leading
order contribution of $K$ for the 1D electron gas model 
with two-particle interaction $U(x)$ is given by \cite{Schoenhammer05} 
\begin{eqnarray}
\label{Kpoor}
K \approx 1- \left[\frac{\tilde U(0)}{2 \pi v_F} -  
\frac{\tilde U(2 k_F)}{2 \pi v_F}\right] = 1- \alpha 
\end{eqnarray}
with the forward- [$\tilde U(0)$] and 
backward-scattering [$\tilde U(2 k_F)$] amplitudes.

In the Appendix we derive the poor man's RG equation 
  (\ref{poormansRGequation}) for the 1D
  electron gas model within the fRG framework. This requires
  approximations which go beyond the ones used to obtain
  Eqs.~(\ref{finalflowsigma}) and (\ref{finalflowgamma}).

\section{A comparison of results for the conductance and the
 renormalized impurity potential}
\label{results}

In this section we compare the results for the conductance 
and the renormalized impurity potential
obtained using the two fermionic RG methods. We start with the two
limits of weak and strong bare impurities.   

\subsection{The conductance for weak and strong impurities}

The solution Eq.~(\ref{poormansRGsol}) of the poor man's RG equation
can easily be expanded in the limits of weak ($|r_0/t_0| \ll 1$) 
and strong ($|t_0/r_0| \ll 1$) bare impurities. In the first case this
gives 
\begin{eqnarray}
\label{poorweak}
|t^s|^2 -1 \approx - \left| \frac{r_0}{t_0} \right|^2 \, \left|
  \frac{s}{\lambda_0} \right|^{-2 \alpha} 
\end{eqnarray}
while in the second one finds
\begin{eqnarray}
\label{poorstrong}
|t^s|^2  \approx  \left| \frac{t_0}{r_0} \right|^2 \, \left|
  \frac{s}{\lambda_0} \right|^{2 \alpha} \; .
\end{eqnarray}
Within the bosonization approach the two relevant scaling exponents
are give by $2(K-1)$ and $2(1/K-1)$ [see Eqs.~(\ref{weaklimit}) and
(\ref{stronglimit})]. With Eq.~(\ref{Kpoor}) one finds 
\begin{eqnarray}
\label{konsi}
2(1/K-1) \approx -2(K-1) \approx 2 \alpha
\end{eqnarray}
to leading order in the two-particle interaction 
and Eqs.~(\ref{poorweak}) and (\ref{poorstrong}) are consistent
with the bosonization results  Eqs.~(\ref{weaklimit}) and
(\ref{stronglimit}) if $s$ is replaced by $T$. One is thus 
tempted to conclude that the poor man's RG result 
Eq.~(\ref{poormansRGsol}) describes the full crossover from the
``perfect chain fixed point'' to the ``open chain fixed point'' 
as long as the interaction remains small $\alpha \ll 1$. 
Below we argue that the analytic Breit-Wigner 
form Eq.~(\ref{poormansRGsol}) is incomplete and that the apparent  
agreement with the bosonization result heavily relies on the subtle 
relation Eq.~(\ref{konsi}) (which can be traced back to what is called
``duality'' in the field-theoretical approach).
 
In the two limits considered in the present subsection also the fRG
flow equations can be tackled analytically provided that the flow of
the effective nearest-neighbor interaction is neglected. In
Ref.~\cite{frg1} it was shown that at $T=0$ for weak bare impurities
\begin{eqnarray}
\label{frgweak}
\Sigma_{k_F,-k_F}^\Lambda \sim \Lambda^{-[\tilde U(0) - \tilde U(2
  k_F)]/(2 \pi v_F)}
\end{eqnarray}
rather independently of the model considered (that is not only for the
lattice model with nearest-neighbor interaction and independent of the
type of impurity considered). Within the Born
approximation of single-particle scattering theory the 
transmission (conductance) can be obtained from the backscattering
amplitude $\Sigma_{k_F,-k_F}^\Lambda$. 
Assuming that $\Lambda$ in Eq.~(\ref{frgweak}) can be replaced by
the scale $s$ (as it is routinely done in the poor man's RG; see above)
the transmission shows the correct
scaling behavior with an exponent which agrees to the bosonization
result to leading order in the two-particle interaction. 
In Ref.~\cite{frg1} it was also shown that at $T=0$, for the
lattice model described above, and for a strong hopping impurity
(weak link) the local spectral function close to the impurity 
shows power-law scaling (as a function of energy) with an exponent
equal to that found for an open chain. The latter agrees to
leading order with the bosonization exponent $1/K-1$. Following
standard arguments (Fermi's Golden Rule) \cite{KaneFisher} one can 
then conclude that the conductance at low energy scales shows
power-law scaling with an exponent which agrees to leading order with the
bosonization result $2(1/K-1)$. 

\subsection{The generalized Breit-Wigner form of the transmission}

The expressions (\ref{LB}) and (\ref{transmission}) for the
transmission (conductance) used at the end of the fRG procedure 
can be brought into a form similar to that of the poor man's RG 
Eq.~(\ref{poormansRGsol}). To be specific we consider a site impurity
at $j_0$ (in the bulk of the interacting wire)
of amplitude $V$. In a first step one runs the fRG 
flow down to $\Lambda=0$ and determines the approximate self-energy 
$\Sigma$. Denoting the single-particle version of the
noninteracting part of the Hamiltonian by $h_0$, the $N \times N$
matrix 
\begin{eqnarray}
\label{effhamdef}
h = h_0 + \Sigma + \Sigma^{\rm lead}
\end{eqnarray} 
forms an effective  single-particle ``Hamiltonian'', which is 
non-hermitian because of the leads contribution 
$\Sigma^{\rm lead}$ Eq.~(\ref{leadpotdef}) . After setting 
the renormalized hopping matrix element 
$\tilde t = -1 + \Sigma_{j_0-1,j_0}$ from site $j_0-1$  to site $j_0$
to zero, that is after cutting the chain in two parts, we define 
an auxiliary Green function 
[resolvent matrix of size $(N-j_0) \times (N-j_0)$]
\begin{eqnarray}
\label{auxgreen}
\tilde {\mathcal G}(z) = \left[z- h(\tilde t=0)  \right]^{-1} \; .
\end{eqnarray}   
Via $\Sigma$ the auxiliary Green function still contains
information about all system parameters, in particular $T$ and
$\delta$: $\tilde {\mathcal G}(z,T,\delta) = \tilde {\mathcal G}(z) $. 
Within the framework of single-particle scattering theory and using 
a standard projection technique \cite{frg3} the effective transmission
can be written as 
\begin{equation} 
\label{ND1}
  |t(\varepsilon,T,\delta)|^2 = \frac{4 \Gamma^2(\varepsilon,T,\delta)}{\left[ 
  \varepsilon - \tilde V(T,\delta) - 
  2 \Omega(\varepsilon,T,\delta) \right]^2 + 4  
  \Gamma^2(\varepsilon,T,\delta)} \; ,
\end{equation}
with
\begin{eqnarray}
{\tilde t}^2 \tilde{\mathcal G}_{j_0-1,j_0-1}(\varepsilon+i0,T,\delta) =
\Omega(\varepsilon,T,\delta)-i\Gamma(\varepsilon,T,\delta)  \; ,
\end{eqnarray}
real functions $\Omega$ and $\Gamma$, and the renormalized energy of
the impurity site $j_0$, $\tilde V(T,\delta) = V +
\Sigma_{j_0,j_0}(T,\delta)$. If we now consider $T=0$ and plug
Eq.~(\ref{ND1}) into the generalized Landauer-B\"uttiker formula
(\ref{LB}) we end up with 
\begin{equation} 
\frac{h}{e^2}  G=|t(0,0,\delta)|^2 = \frac{4 \Gamma^2(0,0,\delta)}{\left[ 
\tilde V(0,\delta) + 2 \Omega(0,0,\delta)\right]^2 +
 4 \Gamma^2(0,0,\delta) } \; .
\label{ND1spec}
\end{equation} 
Using $s=\delta$ as our scaling variable this generalized Breit-Wigner
form is similar to that of Eq.~(\ref{poormansRGsol}) obtained within 
the poor man's RG. Using the energy $\delta$ related to the 
inverse size of the interacting part of the system as an energy scale 
is valid also in the poor man's RG according to the construction of the 
RG equation. The characteristic difference of the two approaches 
is the absence of a scale dependence in the first term of the 
denominator of Eq.~(\ref{poormansRGsol}). This will become 
crucial in the following. 

\begin{figure}[tb]
\begin{center}
\includegraphics[width=0.49\textwidth,clip]{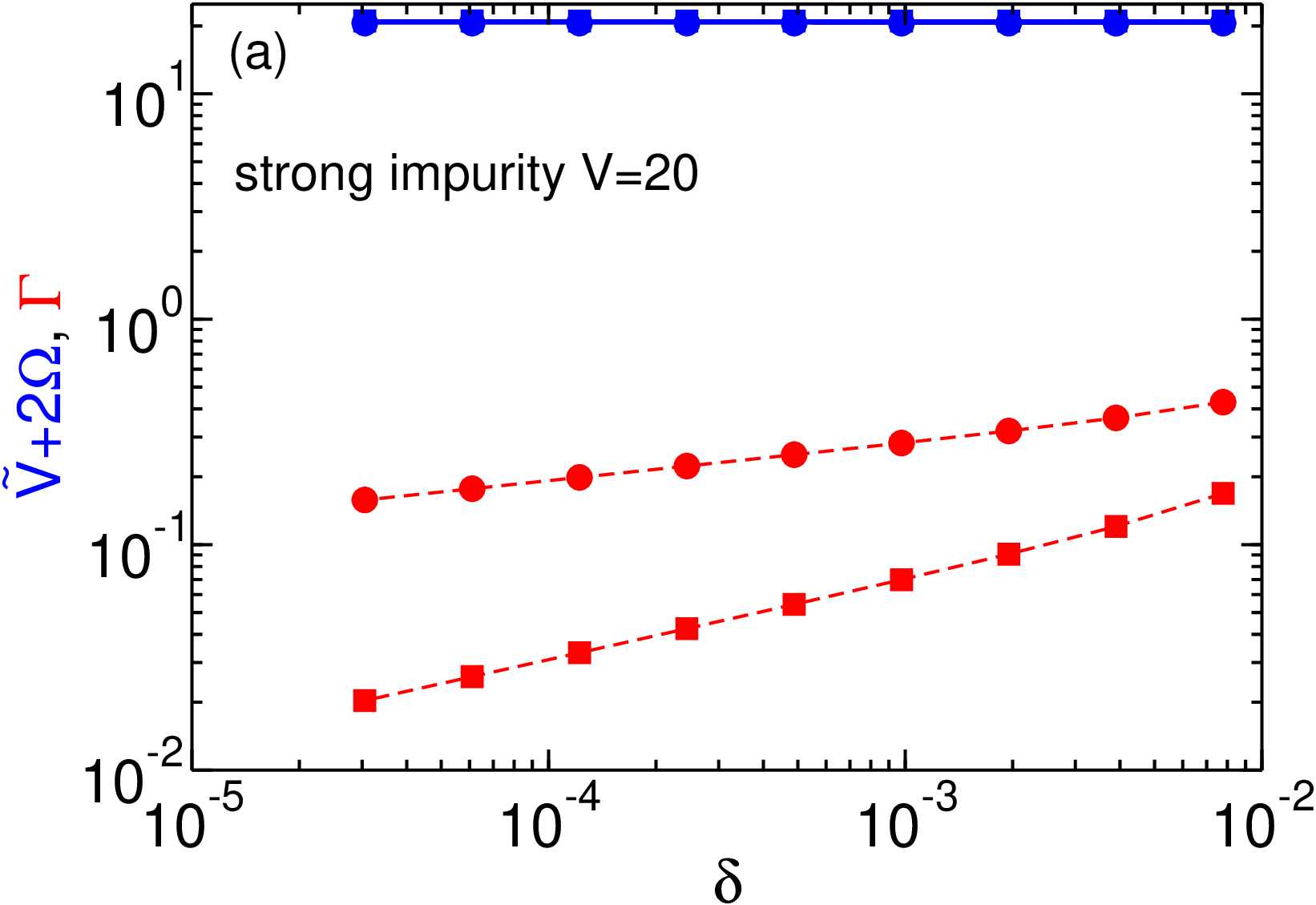}
\includegraphics[width=0.49\textwidth,clip]{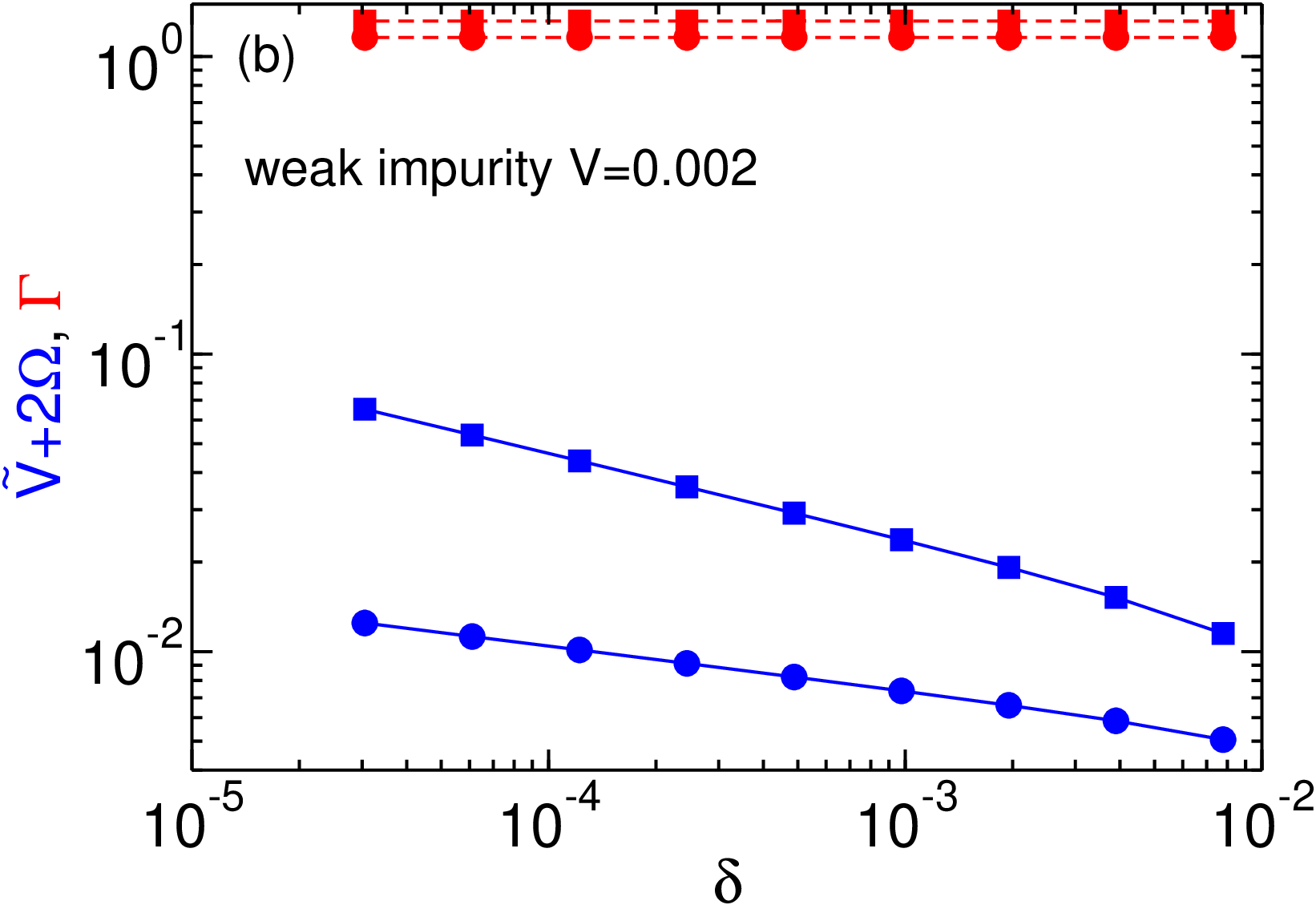}
\end{center}
\caption[]{ $\delta$ dependence of 
$\tilde V+ 2 \Omega$ (solid blue) and $\Gamma$ (dashed red) 
entering the generalized Breit-Wigner form Eq.~(\ref{ND1spec}) at $n=1/2$, 
$U=0.5$ (circles) and $U=1$ (squares) obtained from the truncated fRG. 
In (a) the case of a strong site impurity 
with $V=20$ is shown, while in (b) a weak impurity $V=2 \cdot 10^{-3}$ is considered.  
   \label{fig1}}
\end{figure}

To further develop the relation between the two generalized
Breit-Wigner forms we next analyze the scaling properties of 
$\Omega$ and $\Gamma$. In Fig.~\ref{fig1} results for 
$\tilde V(\delta)+ 2 \Omega(\delta)$ and $\Gamma(\delta)$ obtained by
numerically solving the truncated fRG flow equations for strong
[Fig.~\ref{fig1}(a)] and weak [Fig.~\ref{fig1}(b)] bare impurities are
shown on a log-log scale. The parameters are $U=0.5$ and $1$, $n=1/2$, 
$V=20$ [Fig.~\ref{fig1}(a)] and $V=2 \cdot 10^{-3}$
[Fig.~\ref{fig1}(b)], and the size of the interacting wire varies 
from $N=257$ to $N=65537$. 
For strong bare impurities $\tilde V(\delta)+ 2 \Omega(\delta)$ is
constant, while the imaginary part of the auxiliary Green function shows
power-law scaling as a function of $\delta$. As $\Gamma(\delta) / [\tilde 
V(\delta)+ 2 \Omega(\delta)] \ll 1$ the imaginary part can be neglected in the 
denominator of Eq.~(\ref{ND1spec}). From this one concludes that the exponent 
of $G(\delta)$ is twice as large as that of  $\Gamma(\delta)$.   
The exponent extracted from the fRG data for 
$\Gamma(\delta)$ [see Fig.~\ref{fig1}(a)] agrees to leading 
order in $U$ with the expected result 
$1/K-1$, with $K$ taken from Eq.~(\ref{BetheAnsatz}) \cite{frg3}.
For weak bare impurities $\Gamma(\delta)$ is constant while $\tilde
V(\delta)+ 2 \Omega(\delta)$ follows a power law. Furthermore,
$[\tilde V(\delta)+ 2 \Omega(\delta)] / \Gamma(\delta) \ll 1$ and 
the conductance Eq.~(\ref{ND1spec}) can be expanded in this ratio. 
From this expansion one concludes that the correction of the conductance 
to $e^2/h$ scales with twice the exponent of $\tilde V(\delta)+ 
2 \Omega(\delta)$.  The exponent of $\tilde V(\delta)+ 
2 \Omega(\delta)$ [see Fig.~\ref{fig1}(b)] determined numerically  
indeed agrees to leading order with the expected one $K-1$ \cite{frg3}. 

It is important to note that for $T=0$ Eq.~(\ref{ND1spec}) provides
the {\it exact} analytical form of the conductance even if one goes
beyond the fRG truncation used here. We expect that if
computed exactly $\tilde V(\delta)+ 2 \Omega(\delta)$ and  
$\Gamma(\delta)$ show power-law scaling with the bosonization
exponents (not only the leading order exponents) in the corresponding 
limits of weak and strong impurities.

Identifying $4 \Gamma^2(0,0,\delta)  \leftrightarrow |t_0|^2 |\delta/\lambda_0|^{2 \alpha}$ and 
$[\tilde V(\delta)+ 2 \Omega(\delta)]^2  \leftrightarrow |r_0|^2$ might appear ambiguous as one 
can devide the numerator and denominator of Eq.~(\ref{ND1spec}) by the scale dependent part of
$[\tilde V(\delta)+ 2 \Omega(\delta)]^2 $ and end up with a function similar to    
Eq.~(\ref{poormansRGsol}) in which the first term in the denominator is scale 
independent. However, the above relations become unique if one considers the finite $\varepsilon$ generalization 
of Eq.~(\ref{poormansRGsol}) as given in Eq.~(5) of Ref.~\cite{Nazarov} which must be compared to 
Eq.~(\ref{ND1}). 

These insights provide hints that the
absence of the scale dependence  in the
first term of the denominator of Eq.~(\ref{poormansRGsol}) (the ``real
part'') constitutes a problem for the poor man's RG in the limit of
weak bare impurities. In fact, in the poor man's RG approach 
the ``imaginary part'' carries the scaling in both limits of weak 
and strong impurities. The observation that Eq.~(\ref{poormansRGsol}) 
reproduces the bosonization results in both limits thus relies heavily 
on the fact that $1/K-1 = -(K-1)$ to leading order in the two-particle 
interaction. This is different in the fRG approach, in which two 
different functions carry the scaling properties in the two 
limits. The analysis also reveals that 
for a method that leads to a generalized Breit-Wigner form 
one needs at least two independent RG equations (one for the real and 
one for the imaginary part of $\tilde{\mathcal G}$) to fully describe the
crossover. 

\subsection{The one-parameter scaling function}

We now further elaborate on the differences of the two fermionic RG
approaches by studying the one-parameter scaling properties of the
conductance (transmission).  

That the full crossover from the weak to the strong impurity limit
within the local sine-Gordon model follows a one-parameter scaling 
function was shown analytically in Ref.~\cite{KaneFisher} 
for $K=1/2$. Within this model it was later confirmed for other LL 
parameters \cite{Moon, Fendley} and generalized to microscopic models
in Refs.~\cite{frg2} and \cite{frg3}. The scaling ansatz is given by 
\cite{KaneFisher}
\begin{eqnarray}
\label{oneparascaling}
G(s) = \frac{e^2}{h} \tilde G_K(x) \; , \;\;\;  
x=\left[ s/s_0(U,n,V)\right]^{K-1} \, 
\end{eqnarray}
with the nonuniversal scale $s_0$ and $s$ being $T$ or $\delta$. 
For appropriately
chosen $s_0$ the $G(s)$ curves for different $V$ (but fixed 
$K$) can be collapsed onto the $K$-dependent scaling function $\tilde
G_K(x)$. It has the limiting behavior $\tilde G_K(x) \sim 1-x^2$ for $x
\to 0$ and $\tilde G_K(x) \sim x^{-2/K}$ for $x \to \infty$. For
the following it is important to note that using either
Eq.~(\ref{BetheAnsatz}) or Eq.~(\ref{Kpoor}) the large $x$ 
exponent $-2/K$ has a contribution that is linear in the 
two-particle interaction. It follows from taking the ``open chain
fixed point'' exponent $2(1/K-1)$ (large $x$ behavior) and dividing 
by half the exponent $K-1$ 
of the ``perfect chain fixed point''  which by convention 
was explicitly taken into account in the definition of the variable
$x$ [see Eq.~(\ref{oneparascaling})].

Achieving the form Eq.~(\ref{oneparascaling}) for the result
of the poor man's RG Eq.~(\ref{poormansRGsol}) is 
straightforward. The scaling variable is  $x=|r_0/t_0| 
(s/\lambda_0)^{- \alpha}$ and the scaling function 
\begin{eqnarray}
\label{oneparascalingpm}
\tilde G^{\rm pm}_K(x) = \frac{1}{1+x^2} \; .
\end{eqnarray}
Independent of the strength of the two-particle interaction
it is thus given by the noninteracting scaling function 
$\tilde G_{K=1}$. 
While the behavior at small $x$ is reproduced correctly, for large $x$,
$\tilde G^{\rm pm}_K(x)$ decays $\sim x^{-2}$. 
This can be traced back to the fact that within the
poor man's RG the weak and strong impurity exponents are both 
strictly linear in the two-particle interaction and have the same modulus.  
As mentioned above the correct large $x$ exponent $-2/K$ has a 
contribution linear in the two-particle interaction. 

\begin{figure}[tb]
\begin{center}
\includegraphics[width=0.7\textwidth,clip]{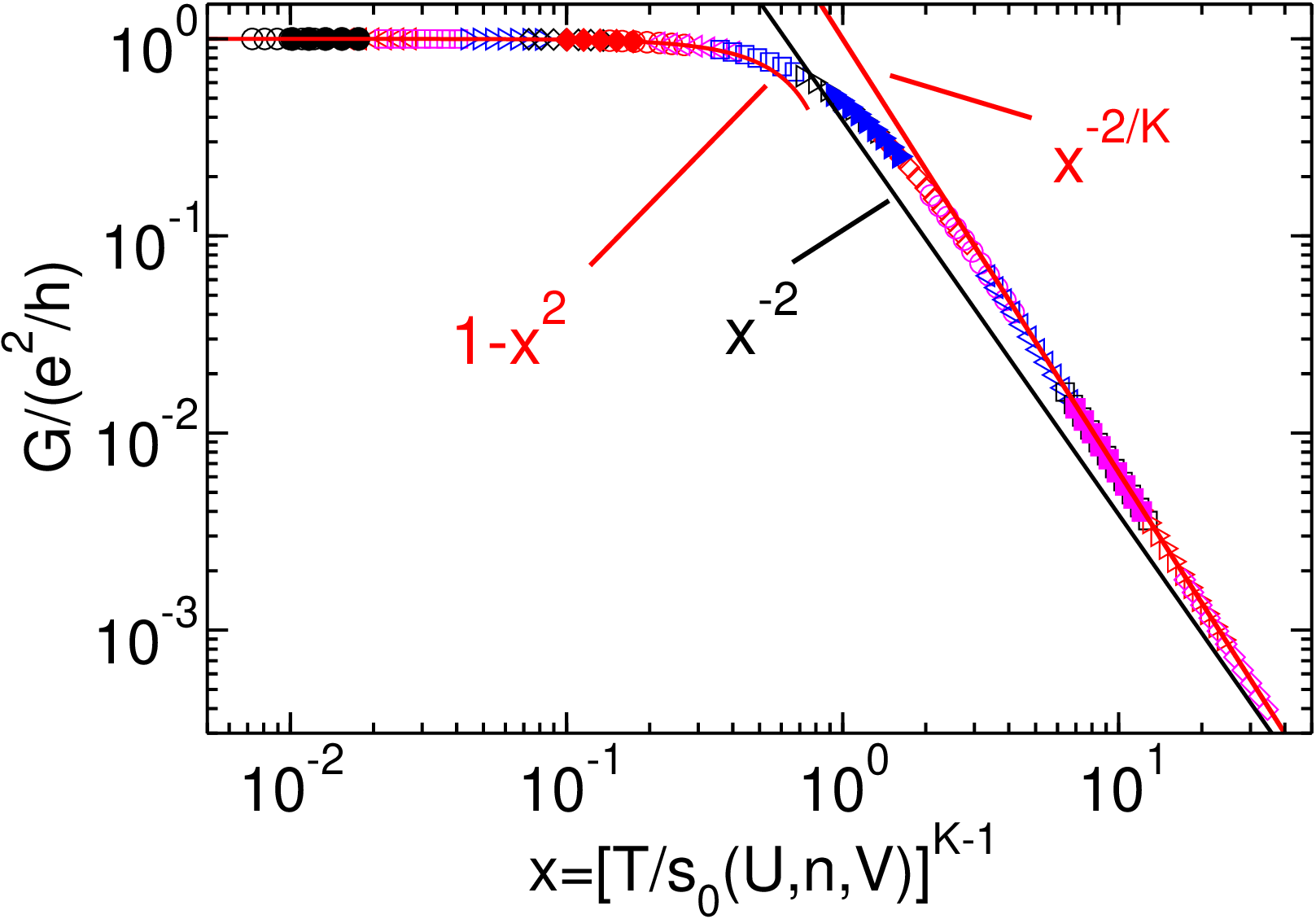}
\end{center}
\caption[]{ One-parameter scaling plot of the conductance obtained by 
the truncated fRG procedure. Open
  symbols represent results obtained for $U=0.5$, $n=1/2$, and
  different $T$ and $V$, while filled symbols were calculated for 
  $U=0.851$, $n=1/4$. Both pairs of $U$ and $n$ lead to the same
  $K^{\rm fRG}-1=-0.15$. The solid red lines indicate the asymptotic 
  behavior for small and large $x$. The solid black line indicates a
  power law with exponent $-2$. \label{fig3}}
\end{figure}

In Fig.~\ref{fig3} we show conductance data obtained
numerically from the fRG procedure at a fixed small 
$\delta$ and for temperatures $ T \gg \delta$. For appropriately chosen 
$s_0(U,n,V)$ the data for different $V$ and $U=0.5$, $n=1/2$ 
(open symbols) fall onto a single curve. The different symbols and colors 
indicate different $V$. 
A similar collapse is found for $T=0$ and $s \to \delta$ \cite{frg2}.  
In the definition of $x$ [see Eq.~(\ref{oneparascaling})] 
we used the fRG approximation for $K-1$ (for  $U=0.5$ and $n=1/2$, 
$K^{\rm fRG}-1=-0.15$). The same $K-1$ can be found at filling 
$n=1/4$ for $U=0.85$. Data for these parameters are shown as 
the filled symbols in Fig.~\ref{fig3}. The collapse confirms 
that the one-parameter scaling function indeed only depends on 
$K$ and not on $U$ and $n$ separately. 

In contrast to $\tilde G^{\rm pm}_K(x)$, 
$\tilde G^{\rm fRG}_K(x)$ {\it depends} on $K$ and for large
$x$ shows a power-law decay with an exponent different from $-2$ 
(see the black line in Fig.~\ref{fig3}). In the fRG approximation the
two exponents $1/K-1$ and $-(K-1)$ are determined numerically from 
the temperature dependence of the conductance for weak 
and strong bare impurities (or, with the same result, as described in
the last subsection). In contrast to the poor man's 
method they have higher (than leading) order corrections 
and the $U$-dependent part does not cancel if the scaling ansatz 
is applied. In fact, as shown in Figs.~5 and 7 of Ref.~\cite{frg3} the
bosonization and fRG exponents agree well beyond the
linear regime in the limits of weak and strong impurities, even though 
formally only the leading order
Taylor coefficients are the same (as terms of order $U^2$ and higher are only
partially kept in the truncated fRG scheme). In the light of these 
observations 
it is interesting to ask whether the fRG approximation of 
the ratio $2(1/K-1)/(K-1) = -2/K$, that is the large $x$ decay 
exponent of the scaling function, has the correct leading order (in
$U$) behavior. Unfortunately, the numerical accuracy of the data for
the two exponents is not high enough to make definite statements about
this. For small $U$ (that is in the linear regime of $-2/K$), 
both exponents are small and any small error (in particular of 
the denominator) will lead to an inaccurate ratio. 
For $K$ close to 1, for 
which the truncated fRG works best, the complete form  of the scaling 
function (beyond the two
limits $x \to 0$ and $x \to \infty$) 
is not known analytically within the local sine-Gordon 
model, and no direct comparison with the fRG approximation is 
possible. But even for $K=1/2$, the largest LL parameter (that is the
smallest two-particle interaction) for 
which  $\tilde G_K(x)$ was computed in the field-theoretical model, 
the fRG data of our lattice model are surprisingly close to 
the curve obtained using bosonization as shown in Fig.~4 
of Ref.~\cite{frg2}. In this respect the fRG method goes significantly 
beyond the poor man's approach.  

\subsection{The decay of the effective, renormalized impurity potential}

It seems to be possible to understand the problems of the poor man's RG
close to the ``perfect chain fixed point'' in terms of the real space 
properties of the effective impurity potential. As described in 
the introduction in the construction of the poor man's RG equation 
the effective impurity potential always decays as the
inverse distance from the impurity, while the prefactor 
is renormalized. This holds up to the spatial scale set by the 
infrared cutoff (e.g.~the temperature $T$) and regardless of the 
strength of the bare impurity. Using the fRG approach we now show 
that this is correct for strong bare impurities  and on 
asymptotically small scales. For weak bare impurities and 
on large to intermediate scales the renormalized impurity potential 
decays more slowly, an observation which can also be related to the results 
known from bosonization.  

\begin{figure}[tb]
\begin{center}
\includegraphics[width=0.7\textwidth,clip]{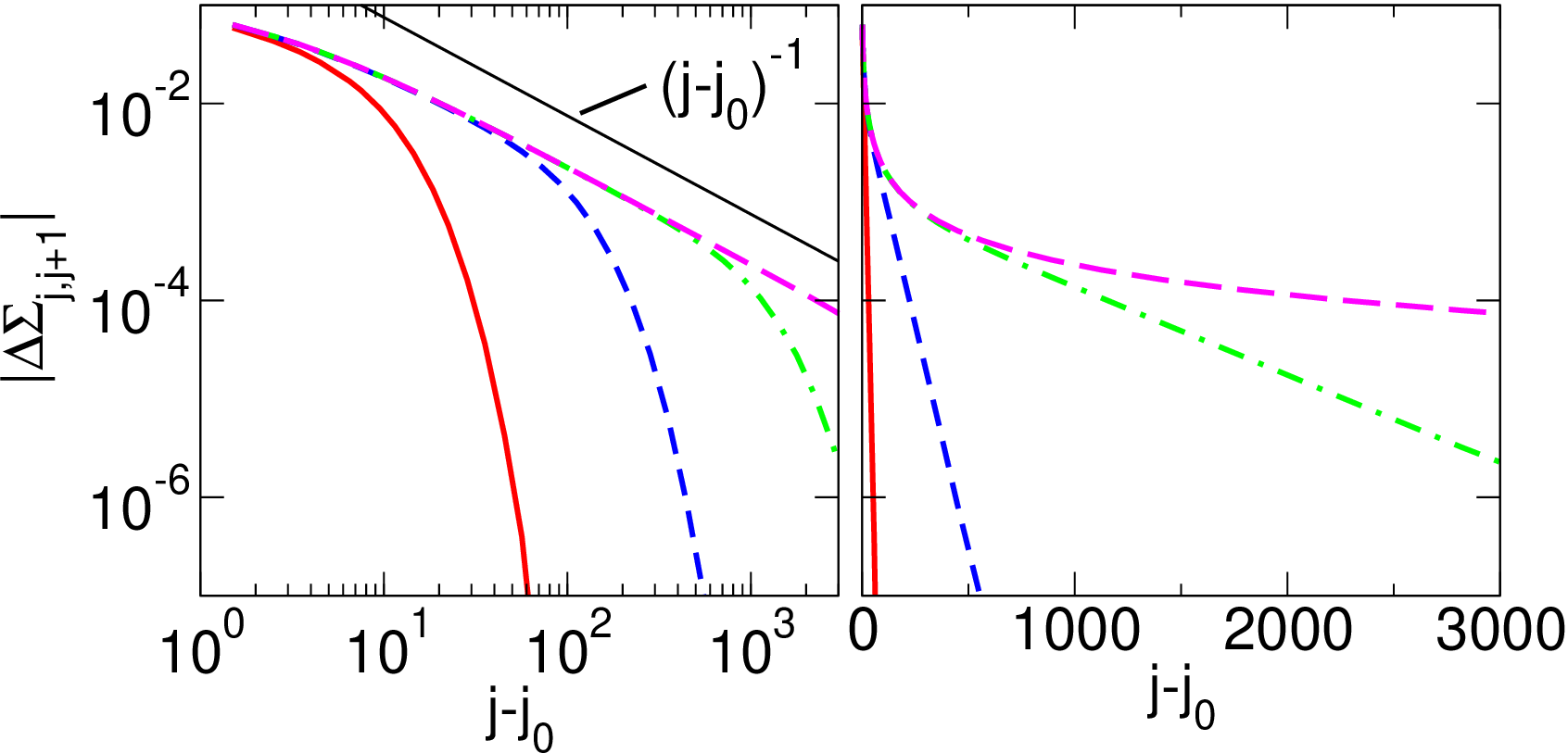}
\end{center}
\caption[]{ Decay of the oscillatory part of the off-diagonal matrix
  element of the self-energy away from a single hopping impurity at
  bond $j_0,j_0+1$. Results for $t'=0.1$, $j_0=5000$, $N=10^4$, $U=1$,
  $n=1/2$ and different temperatures $T=10^{-1}$ (solid red),
  $T=10^{-2}$ (short dashed blue), $T=10^{-3}$ (dashed-dotted green), and
  $T=10^{-4}$ (long dashed magenta) are presented.  The left panel
  shows the data on a log-log scale, the right panel on a linear-log scale.
  For comparison the left panel contains a power law $(j-j_0)^{-1}$
  (thin black line). \label{fig4}}
\end{figure}

In Fig.~\ref{fig4} fRG results for the decay of the oscillatory part 
of $\Sigma_{j,j+1}$ away from $j_0$ for a strong bare impurity 
at different $T$ are shown. Here we used a hopping impurity with 
$t'=0.1$, that is a hopping across the bond $j_0,j_0+1$ which 
is only 10\% of the hoppings across all other bonds. Generically, 
for the lattice model under consideration, both the diagonal and 
off-diagonal parts of $\Sigma$ show decaying oscillations. 
For
symmetry reasons (particle-hole symmetry) $\Sigma_{j,j}=0$ for 
half-filling $n=1/2$ and hopping impurities, as chosen 
in Fig.~\ref{fig4}.  The figure clearly shows that $\Sigma_{j,j+1} \sim
|j-j_0|^{-1}$ up to a scale $\sim T^{-1}$ beyond which the oscillation
decays exponentially. 
For more general parameters and strong bare 
impurities $\Sigma_{j,j} \neq 0$ shows the same characteristics. 
This result is in agreement with the reasoning leading to the 
poor man's RG equation. 

\begin{figure}
\begin{center}
\includegraphics[width=0.6\textwidth,clip]{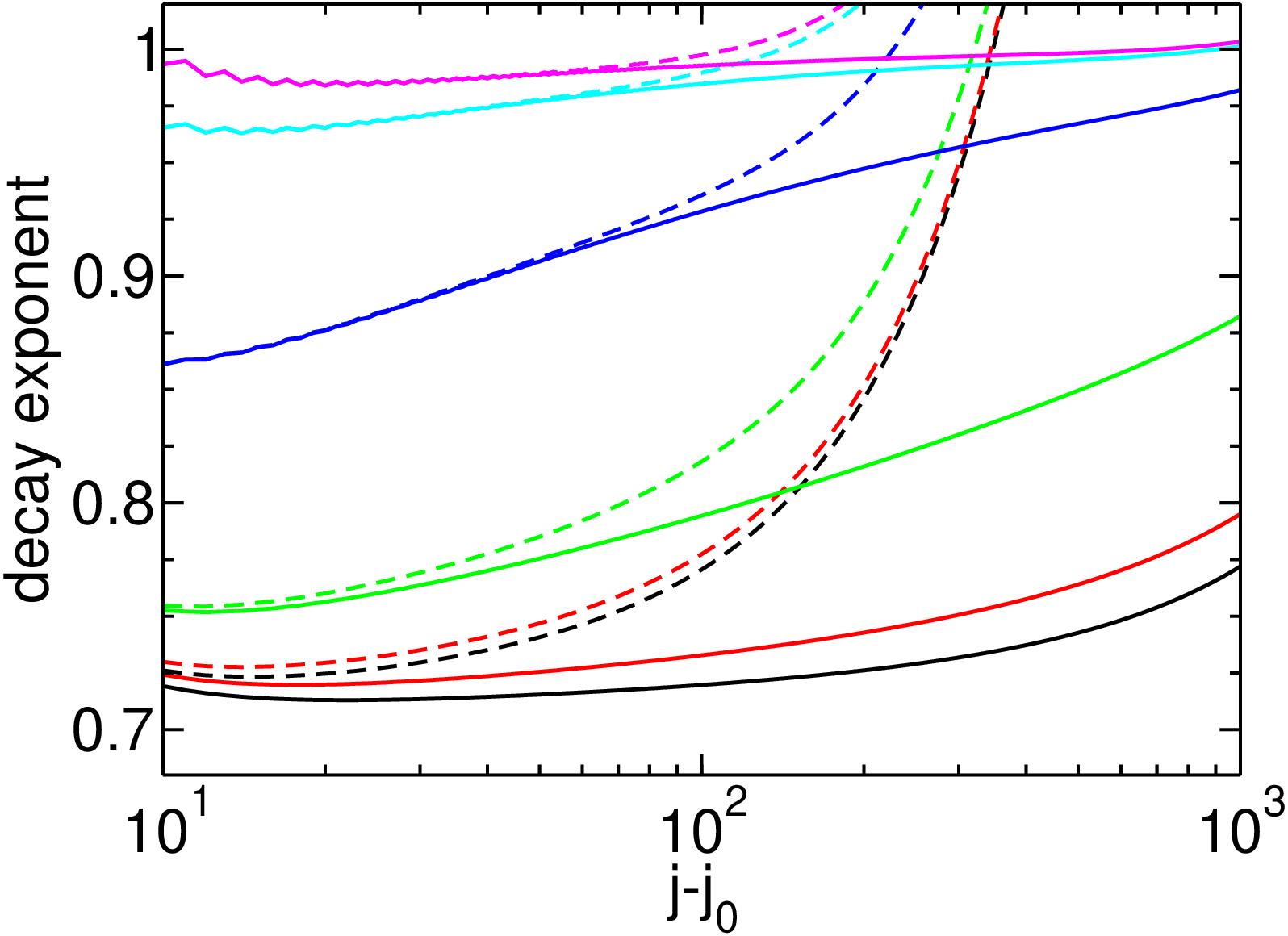}
\end{center}
\caption{Effective exponent for the decay of oscillations of 
 $\Sigma_{j,j}$ as a function of the distance from a site impurity 
 of strengths $V=0.01$, $0.1$, $0.3$, $1$, $3$, $10$ (from bottom 
 to top) at $T=10^{-3}$ (dashed) and $T=10^{-4}$ (solid), 
half-filling $n=1/2$ and interaction 
 strength $U=1$. The impurity is located at the center of a chain with 
 $N=10^5$ sites. \label{fig5}}
\end{figure}

In Fig.~\ref{fig5} we show the effective exponent for 
the decay of the oscillations of $\Sigma_{j,j}$ as a function of 
the distance from a site impurity of different strengths $V$ 
obtained at $T=10^{-3}$ (dashed) and $T=10^{-4}$ (solid). 
For weak bare impurities and on intermediate scales the effective 
impurity potential follows a power-law decay with a $U$-dependent
effective exponent (see the plateau at value $\approx 0.71$ 
for $V=0.01$ at $T=10^{-4}$). 
Only for larger distances and lower temperatures, that 
is on lower-energy scales, or for stronger bare impurities 
this behavior crosses over to the strong impurity limit with 
decay exponent $1$.  This behavior seems not to be included in the
poor man's approach. The steep rise of the effective
exponent on a scale $j-j_0 \approx v_F/T$ indicates the crossover 
to exponential decay (see Fig.~\ref{fig4}). 
The off-diagonal component of the self-energy $\Sigma_{j,j\pm1}$ shows
similar behavior. 

When studied as a function of $U$ the effective exponent at the 
weak impurity plateau is found to be close to $K$. This exponent 
$K$ can be related to the results known from bosonization. 
For weak impurities and on large to intermediate energy scales 
the effective impurity stays small and is approximately 
$\sim \Theta(v_F/T-|j-j_0|) 
\cos{[2 k_F (j-j_0)]}/|j-j_0|^K$. To obtain the reduction of the 
transmission with respect to 1, one can then treat the renormalized 
impurity potential within the Born approximation. This requires 
the $2 k_F$ component of the impurity potential. Fourier 
transforming the above potential leads to $\Sigma^\Lambda_{k_F,-k_F} 
\sim T^{K-1}$ and finally to $G- e^2/h \sim - T^{2(K-1)}$, that is 
the scaling behavior known from the local sine-Gordon model.    

\section{Including the spin degree of freedom}
\label{spin}

Both fermionic RG methods were extended to study inhomogeneous
LLs with spin. Within the generalization of the local sine-Gordon model 
to the case with spin the physics is in many respects similar to 
that of spinless inhomogeneous LL. The system is characterized by 
the ``perfect'' and the ``open chain fixed points'' with scaling dimensions 
$K/2+1/2$ and $1/(2 K)+1/2$, respectively \cite{KaneFisher,Furusaki}. 
We here assume a spin-rotationally invariant interaction. 
From the scaling dimensions the exponents of the power-law corrections 
to the fixed-point conductance and the exponent of the local 
spectral weight follow as in the spinless case. 

To obtain the scaling exponents to first order in the two-particle 
interaction within fermionic RG methods it is essential to keep the 
flow of the static two-particle vertex (the effective interaction). 
As it is known from the so called g-ology analysis for homogeneous LLs 
the backscattering component between electrons with opposite spin 
flows to zero if the infrared cutoff is lowered \cite{Solyom}. 
In the poor man's RG the flow equation for the transmission at momentum
$k_F$ was supplemented by the g-ology RG equations for the static 
two-particle vertex of a homogeneous 
TL model. This way the prefactor $\alpha$ in Eq.~(\ref{poormansRGequation}) 
becomes $\Lambda$-dependent. For strong bare impurities 
a new feature appears. If for the two-particle interaction 
$\tilde U(0)- 2 \tilde U(2k_F) < 0$, that is for repulsive interactions
with a sizable bare backscattering component,  
the conductance across  a single impurity first increases as a
function of the energy scale $s$ before it is suppressed on 
asymptotically small scales. This effect is not captured within the 
local sine-Gordon model. 

\begin{figure}[tb]
\center{\includegraphics[width=0.6\textwidth,clip]{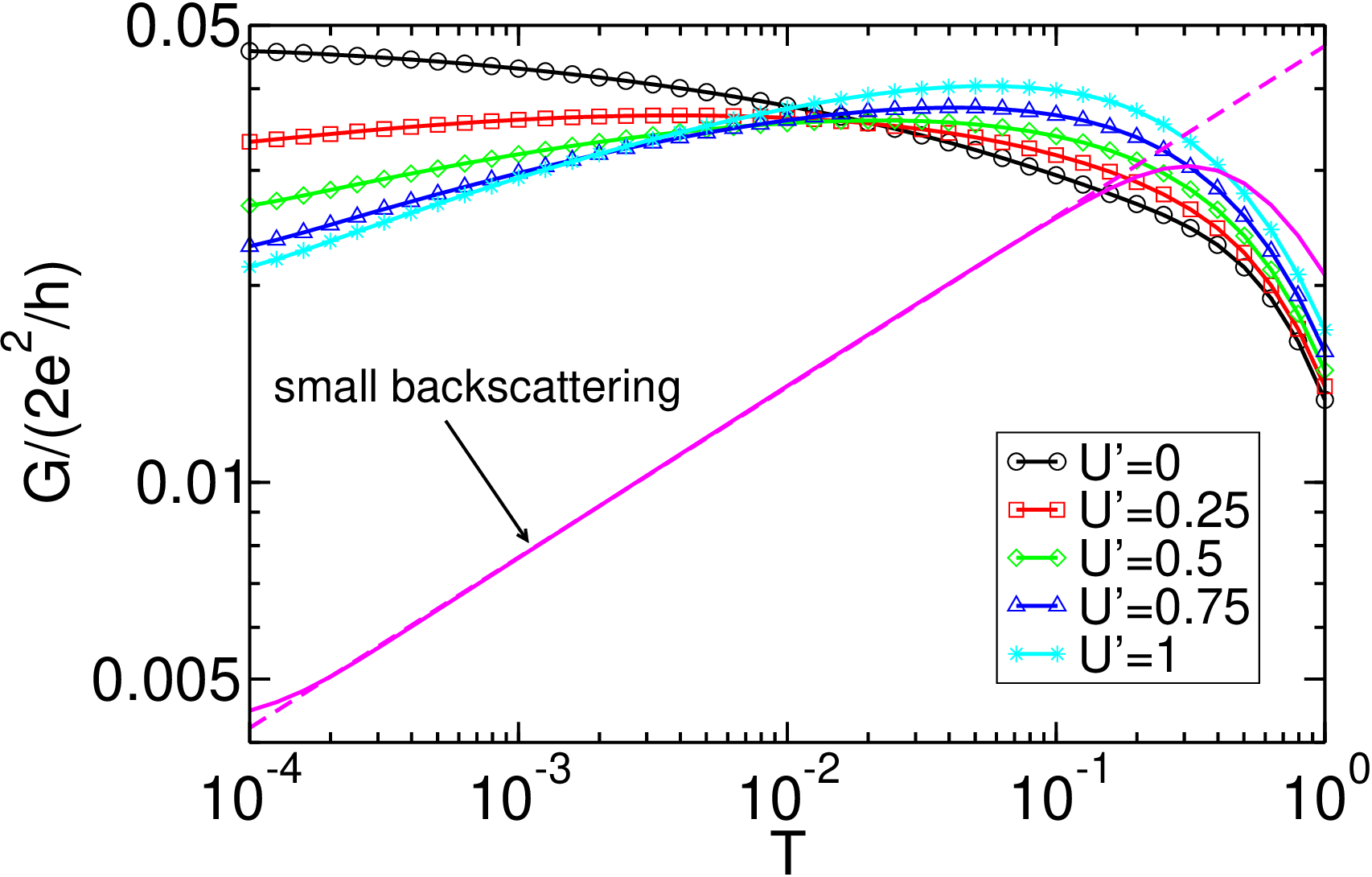}} 
\caption{\label{fig6} 
 Temperature dependence of the conductance for the extended Hubbard
 model with $N = 10^4$ sites and a single site impurity of strength 
 $V = 10$, for a Hubbard interaction $U = 1$ and various choices 
 of the nearest-neighbor interaction $U'$ obtained by fRG. 
 The density is $n=1/2$, except for the lowest curve,
 which has been obtained for $n=3/4$ and $U' = 0.65$ (leading to a 
 very small backscattering interaction).
 The dashed line is a power-law fit for the latter parameter set.}
\end{figure}

The fRG approach was generalized to the extended Hubbard model 
with local interaction $U$ and nearest-neighbor interaction $U'$ \cite{frg5}.
In a wide range of parameters -- to which we stick in the present context -- 
this model shows LL behavior.  The flow equations for the different 
components of the static two-particle vertex go beyond the g-ology 
approximation (see Figs.~2 and 3 of Ref.~\cite{frg5}), but the 
feedback of the inhomogeneity on the flow of the vertex (via the 
self-energy) is neglected.  As shown in 
Fig.~\ref{fig6} also within the fRG approach for a strong site impurity 
$V=10$ the conductance first increases when the temperature is 
lowered before the asymptotic power-law suppression sets in.
For the extended Hubbard model, $\tilde U(0) - 2 \tilde U(2k_F) 
= 2U'[1 - 2\cos(2k_F)] - U$, which can be positive or negative 
for $U,U'>0$, depending on the density and the relative 
strength of the two interaction parameters. 
At quarter-filling $n=1/2$, $\tilde U(0) - 2 \tilde U(2k_F)$ is negative 
and therefore leads to an enhanced conductance for $U' < U/2$.
In case of a  negative $\tilde U(0) - 2 \tilde U(2k_F)$ 
the crossover scale is given by 
\begin{equation}
 T_c \propto \exp \left( 
 \frac{2\pi v_F}{\tilde U(0) - 2 \tilde U(2k_F)} \right) \; ,
\end{equation}
and becomes exponentially small for weak interactions.
For $U'=0$ the conductance increases as a function of 
decreasing $T$ down to the lowest temperatures shown in Fig.~\ref{fig6}. 
For increasing nearest-neighbor interactions $U'$ a suppression
of $G(T)$ at low $T$ becomes visible, but in all the data obtained
at quarter-filling $n=1/2$ it is much less 
pronounced than what one expects from the asymptotic power 
law with exponent $1/K-1$. 
By contrast, the suppression is much stronger and follows the
expected power law more closely if parameters are chosen such
that the bare two-particle backscattering becomes negligible,
as can be seen from the conductance curve for $n=3/4$ and 
$U'=0.65$ (magenta) in Fig.~\ref{fig6}. The value of $K$ \cite{frg5}
for these parameters almost coincides with the one for another
parameter set in the plot, $n=1/2$ and $U'=0.75$ (blue triangles), but the
behavior of $G(T)$ is completely different.
Note that at $T \sim \delta$ the scaling is cut off 
as can be seen at the low $T$ end of the magenta curve in the 
figure. The scaling 
exponents (if accessible, as e.g.~for the magenta curve) 
extracted from the $T$- or $\delta$-dependence of the 
conductance (for weak and strong bare impurities) as well as 
the exponent extracted from the $\omega$-dependence of the 
local spectral function agree to leading order in the two-particle 
interaction with the bosonization results.

\begin{figure}[tb]
\includegraphics[width=0.49\textwidth,clip]{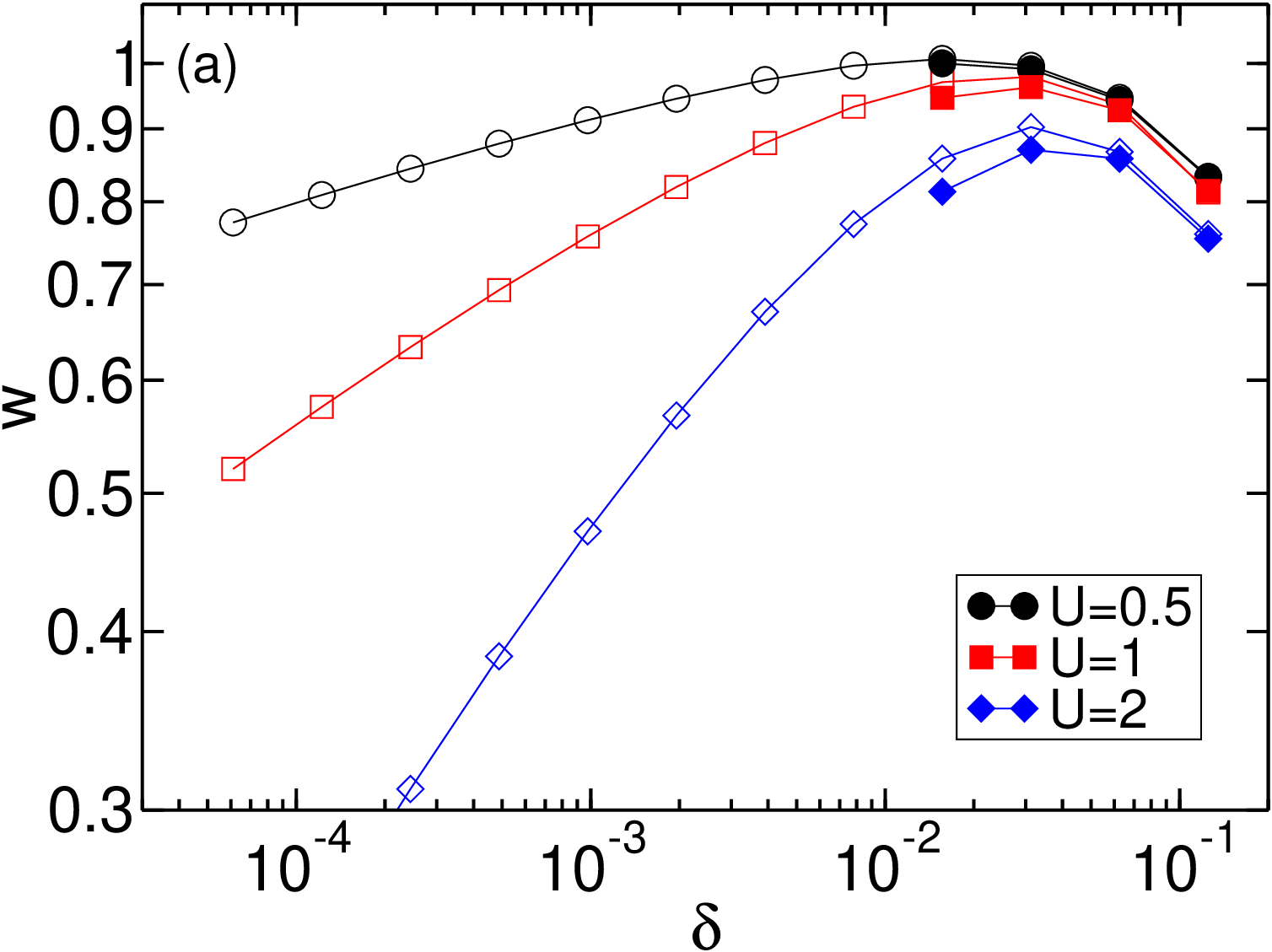}
\includegraphics[width=0.49\textwidth,clip]{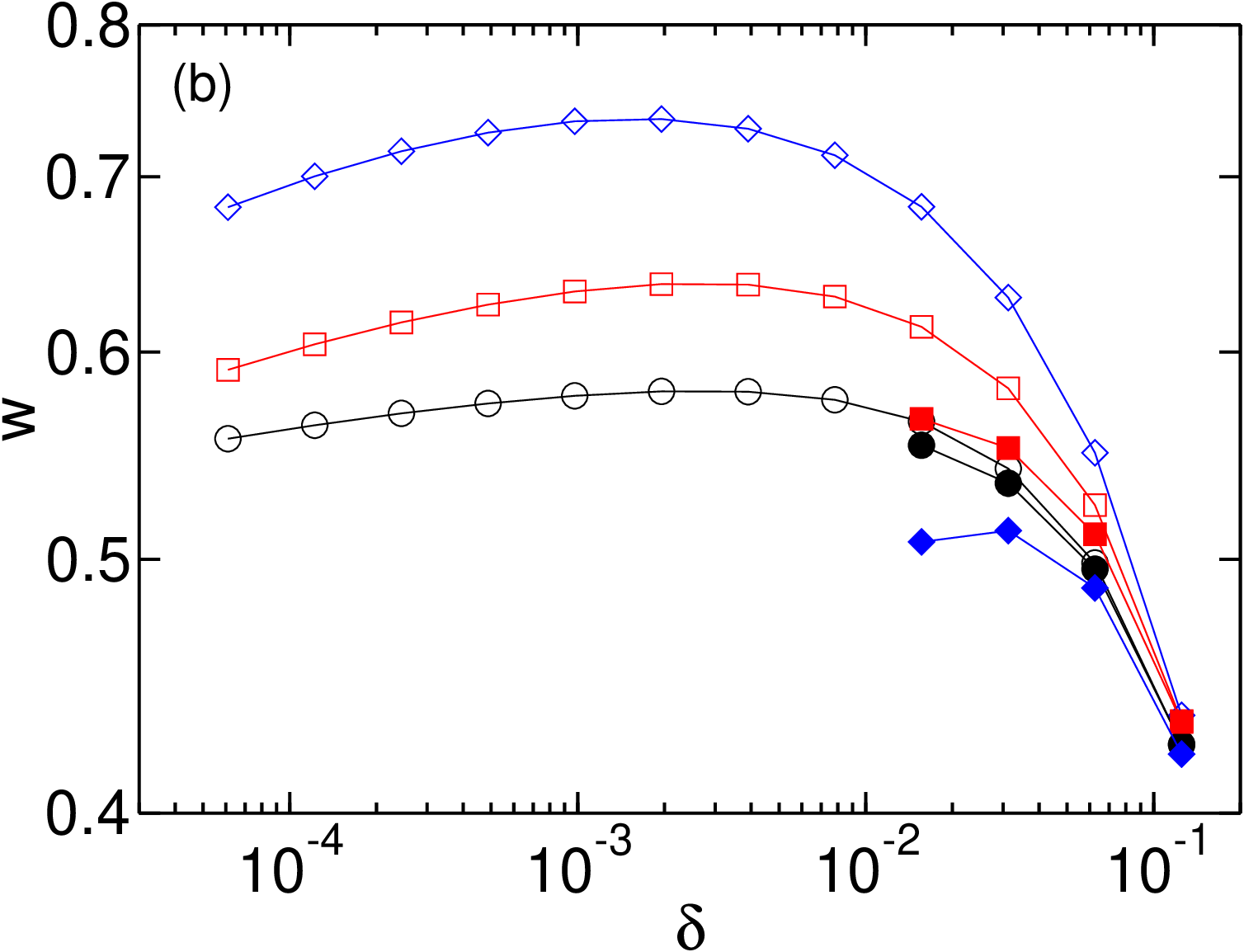}
\caption{\label{fig7} 
 Spectral weight at the Fermi level
 near a hopping impurity $t'=0.5$ as a function of $\delta=v_F/N$
 for the extended Hubbard model with $U' = U/\sqrt{2}$, 
 for various choices of $U$.
 (a) $n = 3/4$ (leading to small backscattering). (b) 
 $n = 1/2$ (leading to sizable backscattering).
 Results from the fRG (open symbols) are compared to DMRG data 
 (filled symbols).}
\end{figure}

Using the fRG we are in a position to even address very subtle
questions such as how accurately the crossover scale 
is reproduced by the truncated procedure. 
To this end we need exact or numerically exact data to compare our 
fRG results with. As no such data are available for the
conductance of the extended Hubbard model with a local impurity
we compute the spectral weight $w$ at $T=0$, close to the impurity, 
and at $\omega=\mu$ as a function of $\delta$ for this model
and compare it to very accurate numerically obtained 
density-matrix renormalization group (DMRG) data \cite{DMRGrev}. 
We here decouple the leads (as they are not included 
in the standard DMRG approach) and study a finite system of length $N$.
In Fig.~\ref{fig7} we present fRG and DMRG results for $w$ taken at site 
$j_0$ for a hopping impurity with $t'=0.5$, $U'=U/\sqrt{2}$, different $U$, 
and $n=3/4$ [Fig.~\ref{fig7}(a); small backscattering] as well as 
$n=1/2$ [Fig.~\ref{fig7}(b); sizable backscattering]. 
The spectral function $\rho_j(\omega)$ close to the impurity site and the 
spectral weight $w(\delta)$ show a similar crossover behavior as $G(T)$ 
[see Fig.~\ref{fig7}(b)].
For parameters at which the bare two-particle backscattering is small 
as in Fig.~\ref{fig7}(a) the fRG and DMRG data agree quite
well up to the smallest $\delta$, that is the largest system sizes $N$ 
that can be studied by DMRG. A similar agreement is found in the spinless
case. 
For $\delta \to 0$, $w(\delta)$ follows a power law  
with an exponent which agrees to leading order in the two-particle 
interaction with the bosonization result $1/(2K)-1/2$. 
In contrast, for parameters 
with a sizable backscattering as in Fig.~\ref{fig7}(b) the DMRG and fRG 
data agree only for sufficiently large $\delta$ (small $N$). For 
larger $U$ and $U'$ the deviation increases and the crossover 
scale of the truncated fRG is easily an order of magnitude too small 
[see the data for $U=2$ in Fig.~\ref{fig7}(b)]. 
As for the conductance the power-law decay sets in only for energies 
sufficiently smaller than the crossover scale. 
We thus conclude that for a quantitative description of the 
crossover behavior specific to the spinful case 
one further has to improve the fRG 
procedure. We expect that including the feedback of the 
self-energy on the flow of the two-particle vertex, that is the 
feedback of the impurity on the flow of the effective two-particle 
interaction, is essential here and will lead to an improved approximation. 
The fRG provides a formal framework to address this question.

\section{Summary} 
\label{summary}

Two fermionic RG methods were developed over the last years 
to investigate the physics of 
inhomogeneous LLs, with a special emphasis put on the transport 
properties of such systems. The poor man's RG (resummation 
of ``leading-log'' divergences) and the fRG (full functional dependence of 
the effective impurity potential) were set up to supplement the 
results obtained using bosonization for an effective field-theoretical 
model (the local sine-Gordon model) in various respects. 
We critically reviewed the results 
obtained within both approaches for the simplest setup, a 1D quantum
wire with two-particle interaction and a single local impurity. 
This system is characterized by two fixed points: the ``perfect chain 
fixed point'' which is unstable for repulsive two-particle interactions
and the ``open chain fixed point'' which is stable. Close to the
fixed points observables show power-law scaling as a function of 
energy variables. 
In the literature both fermionic RG methods are described as 
being applicable to capture the full crossover from one 
to the other fixed point of the single impurity problem 
while keeping interaction effects only to leading order. 

By comparing 
the outcomes of both approaches and relating them to what is known 
from bosonization we collected evidence that the fRG can be 
used to describe the crossover while the poor man's RG seems to have
difficulties close to the ``perfect chain fixed point''. Comparing the 
generalized Breit-Wigner forms in which the conductance in both 
approaches can be written we argued that in the poor man's RG
the derived flow equation for the transmission at $k_F$ must 
be supplemented by at least a second independent equation. For a
single impurity this does not have obvious consequences for 
the scaling behavior of the linear response conductance in the limits
of weak or strong bare impurities because of a subtle relation 
(which can be traced back to what is called ``duality'' in the 
bosonized model) between the scaling exponents of the two fixed
points. For weak bare impurities and on large to intermediate energy
scales the renormalized impurity potential decays with an
exponent $K$ within the fRG (which was argued to be consistent with
what is found in bosonization) while it seems to decay with the
exponent $1$ within the poor man's RG. 
In the fRG approach one obtains a one-parameter scaling function which
depends on the strength of the two-particle interaction while the poor
man's RG leads to the noninteracting scaling function $1/(1+x^2)$.  
These three differences 
can be traced back to the main difference in the construction of 
the two fermionic RG methods. In the poor man's RG only the 
$k_F \to -k_F$ (and vice versa) scattering is kept, while 
the fRG includes the mutual feedback of all scattering channels. 
In fact, for a single impurity 
the poor man's RG equation (\ref{poormansRGequation}) 
can be obtained from the fRG approach by the following steps:
(i) One neglects the flow of the two-particle vertex. 
(ii) One considers scattering states instead of Wannier states. 
(iii) The plane wave matrix elements of the two-particle interaction
are assumed to be constant (up to the g-ology classification). 
(iv) In the overlaps of the plane wave and scattering states one
assumes the asymptotic form of the scattering states even close to the 
local impurity, that is the nontrivial form of the scattering states 
generated during the RG flow is neglected.  This last step leads 
to the decoupling of the scattering channels. Details on this are
  given in the Appendix.

Both approaches can be 
extended to study more complex situations such as junctions of 
several wires, to characterize the corresponding low-energy 
physics. In this respect the bosonization approach is less flexible. 
It is interesting to address the question how both approaches compare 
in situations with a richer fixed-point structure in which the scaling
exponents are not related by a simple transformation (such as $K
\leftrightarrow 1/K$ for the single impurity case). 
A more complex behavior is found for Y-junctions pierced by a magnetic 
flux as studied in Refs.~\cite{Claudio} using field-theoretical 
methods, in Ref.~\cite{Lal} applying the poor man's RG, and in 
Ref.~\cite{frg4a} by fRG. 
In the field-theoretical approach it was possible to identify one of 
the several
fixed points found using the fermionic RG methods and determine its scaling 
dimension as a function of the LL parameter $K$. This fixed point is 
characterized 
by a maximal breaking of time-reversal symmetry, that is the transmission 
probability from leg $i$ to leg $j$ (with $i,j=1,2,3$) of the
Y-junction is unity if $i>j$ and zero if $j>i$ (or vice versa). 
For this fixed point the scaling 
dimension of both fermionic RG methods agree to leading order with the 
one found 
for the field-theoretical model. This has to be contrasted to the
situation at  
another fixed point, which was so far not identified in the field-theoretical 
approach. Both fermionic approaches predict that at this fixed point the 
transmission probability between all legs is equal to $4/9$ 
(``perfect junction 
fixed point''), while 
the scaling exponents obtained differ even to leading order in the 
two-particle interaction. This situation might present an example in which 
the discussed problems of the poor man's RG have a more severe effect
than in the case of a single impurity. To obtain more insights on this 
it would be very 
desirable to identify and characterize the ``perfect junction 
fixed point'' of a Y-junction using a method which does not require 
approximations in the strength 
of the two-particle interaction.

In a very recent preprint Aristov and W\"olfle \cite{Aristov} derive a
RG flow equation for the conductance of a LL with a single impurity
by keeping more than the ``leading-log'' contributions. To describe 
the underlying homogeneous LL the TL model is used (as in bosonization). 
This approach seems to overcome the problems of the poor man's RG and 
in addition can be used for arbitrary amplitudes of the TL two-particle 
interaction. It would be very
interesting to see if this method can be extended to treat systems with
spin and with more complex inhomogeneities.

{\bf Note added in proof.}
After submission of the present manuscript we were informed by 
D.~Sen that he redid the poor man's RG calculation \cite{Lal} for 
the Y-junction. He now obtains a scaling exponent of the
``perfect junction fixed point'' which is consistent with the 
result derived using fRG \cite{frg4a}.

\ack
We thank W.~Metzner and U.~Schollw\"ock for the very fruitful 
collaboration on many of the issues discussed here and S.~Jakobs for very helpful 
comments on the present paper.  This work 
was supported by the DFG through FOR 723 (VM, KS, and HS) and by 
the Alexander von Humboldt foundation through a Feodor Lynen 
fellowship (TE). 

\appendix

\section{Deriving the poor man's RG equation from the fRG}

The poor man's RG equation (\ref{poormansRGequation}) 
can be derived within the fRG framework. To achieve
this we slightly modify the steps which led to 
Eq.~(\ref{finalflowsigma}). For the interacting wire we consider 
the spinless 1D electron gas model instead of a lattice model. Furthermore 
we do not integrate out the ``adiabatically'' connected  
noninteracting (1D electron gas) leads
before setting up the fRG procedure. As a consequence we have to deal with
infinite instead of finite matrices. If the approximations which led
to Eq.~(\ref{finalflowsigma}) are applied this does not change the
formal structure of this flow equation for the frequency independent 
self-energy. We now neglect the flow of the two-particle vertex
altogether and replace $\Gamma^\Lambda$ by the antisymmetrized
interaction $\bar v$. The only energy scale which appears
in the poor man's RG is the cutoff $\Lambda$. In the fRG we therefore consider
the $T=0$ limit and later introduce temperature by stopping the RG flow
at $\Lambda=T$. After these steps the approximate 
flow equation for the self-energy in the momentum state (plane wave) 
basis reads 
\begin{eqnarray}
 \frac{d}{d\Lambda} \Sigma^{\Lambda}_{k',k} =
 - \frac{1}{2\pi} \sum_{\omega = \pm\Lambda} \int dq \, dq' \,
 e^{i\omega 0^+} \, {\mathcal G}^{\Lambda}_{q,q'}(i\omega) \,
 \bar v_{k',q';k,q}  \; ,
\label{finalflowsigma_app}
\end{eqnarray}
with the plane wave matrix elements $ \bar v_{k',q';k,q} $ of the
antisymmetrized two-particle interaction. Within the static
approximation $ {\mathcal G}^{\Lambda}(i\omega)$ can be
interpreted as the resolvent $[i \omega - h_0 - \Sigma^\Lambda]^{-1}$ 
of an effective single-particle problem characterized by the
single-particle version of the noninteracting Hamiltonian $h_0$ and
the flowing scattering potential $\Sigma^\Lambda$. We emphasize that 
the effective potential is restricted to the finite interacting part
of the system. At fixed $\Lambda$ we can therefore define scattering
states $\left| k, + \right>_\Lambda$,  
with $h^\Lambda \left| k, + \right>_\Lambda = \varepsilon_k \left| k + 
\right>_\Lambda$ and $h^\Lambda = h_0 + \Sigma^\Lambda$. 
For large $|x|$ the scattering states are given by ($k>0$)
\begin{eqnarray}
\left< x \right. \left| k, + \right>_\Lambda = \frac{1}{\sqrt{2 \pi}}
\times \left\{ \begin{array}{ll} 
 e^{ikx} + r_k^\Lambda e^{-ikx} & \mbox{for} \; x \to - \infty\\
t_k^\Lambda e^{ikx} & \mbox{for} \; x \to \infty \; .
\end{array} \right.
\label{asymform} 
\end{eqnarray} 
A corresponding expression holds for $k<0$. The scattering states can be
expressed in terms of the plane wave states using the
$T$-matrix \cite{Taylor}
\begin{eqnarray} 
\left| k, + \right>_\Lambda = \left[ 1+ \frac{1}{\varepsilon_k - h_0 + i0}
  T^\Lambda(\varepsilon_k+i0)  \right] \left| k\right> \; .
\label{Tmatrixdef} 
\end{eqnarray} 
The transmission $t_k^\Lambda$ and reflection $r_k^\Lambda$ are given
as plane wave matrix elements of the $T$-matrix \cite{Taylor}
\begin{eqnarray} 
t_k^\Lambda & = & 1 - \frac{2 \pi i}{|v_k|} \left< k \right|
T^\Lambda(\varepsilon_k+i0) \left| k \right> \; , \nonumber \\
r_k^\Lambda & = & - \frac{2 \pi i}{|v_k|} \left< -k \right|
T^\Lambda(\varepsilon_k+i0) \left| k \right> \; ,
\label{trfromT} 
\end{eqnarray} 
with the velocity $v_k = d \varepsilon_k / dk$. 
Using the distorted wave Born approximation \cite{Taylor} 
one can show that a change of the
scattering potential $\Sigma^\Lambda \to \Sigma^\Lambda + d
\Sigma^\Lambda$ leads to the changes 
\begin{eqnarray} 
d t_k^\Lambda & = &  - \frac{2 \pi i}{|v_k|}  \; \mbox{}_\Lambda\!\left< k,- \right| 
d\Sigma^\Lambda \left| k,+ \right>_\Lambda \; , \nonumber \\
d r_k^\Lambda & = &  - \frac{2 \pi i}{|v_k|}  \; \mbox{}_\Lambda\!\left< -k,- \right| 
d\Sigma^\Lambda \left| k,+ \right>_\Lambda  
\label{trchanges0} 
\end{eqnarray} 
of the transmission and reflection amplitudes. With the relation
\begin{eqnarray} 
 \left| k,- \right>_\Lambda & = & t_k^{\Lambda \; *}   \left| k,+ \right>_\Lambda 
+  r_{-k}^{\Lambda \; *} \left| - k,+ \right>_\Lambda 
\label{trchanges0rel} 
\end{eqnarray} 
between the scattering states one obtains
\begin{eqnarray} 
d t_k^\Lambda & = &  - \frac{2 \pi i}{|v_k|} \left( t_k^\Lambda 
\; \mbox{}_\Lambda\!\left< k,+ \right| d\Sigma^\Lambda \left| k,+ \right>_\Lambda + r_{-k}^\Lambda
\; \mbox{}_\Lambda\!\left< -k,+ \right| d\Sigma^\Lambda \left| k,+
\right>_\Lambda  \right) \; ,
\nonumber \\
d r_k^\Lambda & = &  - \frac{2 \pi i}{|v_k|} \left( r_k^\Lambda 
\; \mbox{}_\Lambda\!\left< k,+ \right| d\Sigma^\Lambda \left| k,+ \right>_\Lambda + t_{-k}^\Lambda
\; \mbox{}_\Lambda\!\left< -k,+ \right| d\Sigma^\Lambda \left| k,+ \right>_\Lambda  \right) \; . 
\label{trchanges} 
\end{eqnarray} 
We note in passing that these expressions also hold for lattice
models. 
Combining Eqs.~(\ref{finalflowsigma_app}) and (\ref{trchanges}) 
and transforming into the scattering state basis gives 
\begin{eqnarray} 
\frac{d t_k^\Lambda}{d\Lambda} &=& \frac{i}{|v_k|}   \sum_{\omega =
  \pm\Lambda} \int dq \, \frac{1}{i \omega - \varepsilon_q} 
\left( t_k^\Lambda \; \bar v^{s,\Lambda}_{k,q;k,q} + r_{-k}^\Lambda
  \; \bar
  v^{s,\Lambda}_{-k,q;k,q}   \right) \; ,\nonumber \\
\frac{d r_k^\Lambda}{d\Lambda} &=& \frac{i}{|v_k|}   \sum_{\omega =
  \pm\Lambda} \int dq \, \frac{1}{i \omega - \varepsilon_q} 
\left( r_k^\Lambda \; \bar v^{s,\Lambda}_{k,q;k,q} + t_{-k}^\Lambda
  \; \bar
  v^{s,\Lambda}_{-k,q;k,q}   \right) \; ,
\label{trflow} 
\end{eqnarray} 
with the $\Lambda$-dependent scattering wave basis matrix element 
\begin{eqnarray} 
\mbox{} \hspace{-1cm}\bar v^{s,\Lambda}_{k',l';k,l} = \int dq \, dq' \, dp \, dp' \, 
\left< k',+ \right. \left| q' \right> \left< l',+ \right. \left| p' \right>
\bar v_{q',p';q,p} \left< q \right. \left| k,+ \right> 
 \left< p \right. \left| l,+ \right> 
\label{vsdef} 
\end{eqnarray} 
of the antisymmetrized two-particle interaction. 
In the following we focus on the first equation in (\ref{trflow}).  
 
To compute $\bar v^{s,\Lambda}_{k',l';k,l}$ we implement further 
approximations. In a first step we project the plane wave matrix
element $\bar v_{q',p';q,p}$ on the two Fermi points $\pm k_F$, that
is we consider the g-ology model \cite{Solyom} with
\begin{eqnarray} 
 \bar v_{q',p';q,p} & = & - v_F \alpha  \; \delta(q'+p'-q-p)
 \nonumber \\ && \times
\left\{ \begin{array}{ll} 1  & \mbox{if} \; \mbox{sign} \; q' = \;
    \mbox{sign} \; q  = - \;
    \mbox{sign} \; p' =   - \;
    \mbox{sign} \; p \\
-1  & \mbox{if} \; \mbox{sign} \; q' = \;
    \mbox{sign} \; p  = - \;
    \mbox{sign} \; p' =   - \;
    \mbox{sign} \; q \\
0 & \mbox{otherwise}
\end{array}
\right. 
\label{gologydef} 
\end{eqnarray} 
and $\alpha$ as defined in Sec.~\ref{intro}. The intra-branch
interaction (usually called $g_4$ in the g-ology notation) only leads
to a renormalization of the Fermi velocity. With respect to the
exponent $\alpha$ of the power-law scaling this is an effect of higher
than linear order in the interaction and can thus be
neglected. Secondly and most severly, in computing the overlaps 
of plane wave and scattering states we take the asymptotic 
form Eq.~(\ref{asymform}) of the latter for all $x \lessgtr 0$. 
We thus neglect the nontrivial structure of the 
scattering states within the 
interacting region. The largest contributions to the integrals 
in Eq.~(\ref{vsdef}) follow from momenta close to $\pm k_F$. 
In $\left< k,+ \right. \left| q \right>$ terms with denominators 
$k \pm q +i0$ appear. Depending on the signs of $k$ and $q$ as 
a third approximation we neglect those terms in  which the sum 
or difference of the momenta becomes of order $k_F$ while the 
terms with small denominators are kept. For $k,q>0$ this gives
\begin{eqnarray} 
\left< k,+ \right. \left| q \right> & \approx & \frac{i}{2 \pi} \left(
 \frac{1}{k-q+i0} + \frac{t_k^{\Lambda \; *}}{-k+q+i0}  \right)
\nonumber \\
\left< k,+ \right. \left| -q \right> & \approx & \frac{i}{2 \pi} 
\frac{r_k^{\Lambda \; *}}{-k+q+i0} 
\nonumber \\
\left< -k,+ \right. \left| q \right> & \approx & \frac{i}{2 \pi} 
\frac{r_{-k}^{\Lambda \; *}}{-k+q+i0} 
\nonumber \\
\left< -k,+ \right. \left| -q \right> & \approx & \frac{i}{2 \pi} \left(
 \frac{1}{k-q+i0} + \frac{t_{-k}^{\Lambda \; *}}{-k+q+i0}  \right) \; .
\label{overlaps} 
\end{eqnarray} 
As a last approximation we linearize the dispersion
$\varepsilon_k=\pm v_F(k \mp k_F)$ appearing in Eq.~(\ref{trflow}). 

Under these assumptions a lengthy but straightforward calculation 
yields 
\begin{eqnarray} 
\frac{d t_k^\Lambda}{d\Lambda} = - \frac{v_F \alpha}{v_k} \,
t_k^\Lambda (1-|t_k^\Lambda|^2) \, \frac{1}{\Lambda -i v_F(k-k_F)} \;.
\label{approxflowk} 
\end{eqnarray} 
Remarkably, the different scattering channels are now decoupled, that
is in computing the transmission at momentum $k$ only the same $k$
enters on the right hand side of the flow equation. At the Fermi
momentum one obtains
\begin{eqnarray} 
\frac{d t_{k_F}^\Lambda}{d\Lambda} = - \alpha \,
t_{k_F}^\Lambda (1-|t_{k_F}^\Lambda|^2) \, \frac{1}{\Lambda} 
\label{approxflowkF} 
\end{eqnarray} 
or 
\begin{eqnarray}
\label{poormansRGequationnew}
\frac{d t_{k_F}^\Lambda}{d \ln{(\Lambda/[v_F/d])}} = \alpha
t_{k_F}^\Lambda (1- \left| t_{k_F}^\Lambda \right|^2)
\end{eqnarray}
which is the same as the poor man's RG equation (\ref{poormansRGequation}).


\section*{References}
\begin{thebibliography}{99}
\bibitem{Schoenhammer05}
For a recent review see
Sch\"onhammer K (2005) in {\it Interacting Electrons in Low 
 Dimensions,} Eds.: Baeriswyl D and Degiorgi L, Kluwer 
Academic Publishers
\bibitem{LutherPeschel} Luther A and Peschel I (1974) 
{\it Phys.\ Rev.} B {\bf 9} 2911
\bibitem{Mattis} Mattis D C (1974) {\it J.\ Math.\ Phys.} {\bf 15} 609 
\bibitem{ApelRice} Apel W and Rice T M (1982) {\it 
Phys.\ Rev.\ } B {\bf 26} 7063 
\bibitem{Giamarchi} Giamarchi T and 
Schulz H J (1988) {\it Phys.\ Rev.\ } B  {\bf 37} 325
\bibitem{Herbert} For an introduction see von Delft J and 
Schoeller H (1998) {\it Ann.\ Phys.\ (Leipzig)} {\bf 7} 225 
\bibitem{KaneFisher} 
Kane C L and Fisher M P A (1992) {\it Phys.~Rev.~Lett.} {\bf 68} 1220; 
{\it Phys.~Rev.} B {\bf 46} 7268; {\it Phys.\ Rev.} B {\bf 46} 15233
\bibitem{Furusaki} Furusaki A and Nagaosa N (1993) {\it Phys.~Rev.} B {\bf
    47} 4631
\bibitem{footnote} Here we are interested in 1D quantum wires coupled
  to Fermi liquid leads via reflection-free contacts. In this case the
  conductance in the absence of impurities is given by $G=e^2/h$ as
  discussed in: Safi I and Schulz H J (1995) {\it Phys.~Rev.} B 
  {\bf 52} R17040; Maslov D L and Stone M (1995) {\it Phys.~Rev.} B 
  {\bf 52} R5539; Ponomarenko V V (1995) {\it Phys.~Rev.} B {\bf 52} 
  R8666;  Janzen K, Meden V, and Sch\"onhammer K (2006) {\it
    Phys.~Rev.} B {\bf 74} 085301 
\bibitem{Moon} Moon K, Yi H, Kane C L, Girvin S M, and
  Fisher M P A (1993) {\it Phys. Rev. Lett.} {\bf 71} 4381 
\bibitem{Fendley} Fendley P,  Ludwig A W W, and Saleur H (1995) 
{\it Phys. Rev. Lett.} {\bf 74} 3005 
\bibitem{poor} Matveev K A, Yue D, and Glazman L I (1993)
{\it Phys.\ Rev.\ Lett.} {\bf 71} 3351; 
Yue D,  Glazman L I, and Matveev K A (1994) {\it 
Phys. Rev.} B {\bf 49} 1966 
\bibitem{Nazarov} Nazarov Y V and Glazman L I (2003) 
{\it Phys.~Rev.~Lett.} {\bf
  91} 126804 
\bibitem{Polyakov} Polyakov D G and Gornyi I V (2003) {\it Phys.~Rev.}
  B {\bf 68} 035421
\bibitem{Lal} Lal S,  Rao S, and Sen D (2002) {\it Phys.\ Rev.} 
B {\bf 66} 165327 
\bibitem{Doucot} Kazymyrenko K and Dou\c{c}ot B (2005) {\it Phys.\ Rev.} 
B {\bf 71} 075110 
\bibitem{Belzig} Titov M, M\"uller M, and Belzig W (2006) 
{\it Phys.~Rev.~Lett.} {\bf 97} 237006 
\bibitem{SalmhoferHonerkamp} Salmhofer M and Honerkamp C (2001) 
{\it Prog.\ Theor.\ Phys.} {\bf 105} 1
\bibitem{lecturenotes} Meden V, lecture notes on the ``Functional
  renormalization group'',
  http://www.theorie.physik.uni-goettingen.de/$\sim$meden/funRG/
\bibitem{frg1} Meden V, Metzner W,  Schollw\"ock U, and 
Sch\"on\-hammer K (2002) {\it 
J.\ of Low Temp.\ Physics} {\bf 126} 1147
\bibitem{frg2}  Meden V, Andergassen S,  Metzner W, 
  Schollw\"ock U, and Sch\"onhammer K (2003) {\it 
Europhys.~Lett.} {\bf 64} 769 
\bibitem{frg3}  Enss T, Meden V, Andergassen S, 
Barnab\'e-Th\'eriault X, Metzner W, and Sch\"onhammer K (2005)  
{\it Phys.~Rev.} B {\bf 71} 155401 
\bibitem{frg3a} Jakobs S, Meden V, Schoeller H, and Enss T (2007) 
{\it Phys.~Rev.} B {\bf 75} 035126 
\bibitem{frg4a}
Barnab\'e-Th\'eriault X, Sedeki A, Meden V, and Sch\"onhammer K (2005)
{\it Phys.~Rev.~Lett.} {\bf 94} 136405
\bibitem{frg4b} Barnab\'e-Th\'eriault X, Sedeki A, Meden V, and
  Sch\"onhammer K (2005) {\it Phys.~Rev.} B {\bf 71} 205327 
\bibitem{frg5} Andergassen S, Enss T, Meden V, Metzner W, Schollw\"ock
  U, and Sch\"onhammer K (2006) {\it Phys.~Rev.} B {\bf 73} 045125
\bibitem{Haldane} Haldane F D M (1980) {\it Phys.~Rev.~Lett.} {\bf 45}
  1358
\bibitem{frg6} Andergassen S, Enss T, Meden V, Metzner W, Schollw\"ock
  U, and Sch\"onhammer K (2004) {\it Phys.~Rev.} B {\bf 70} 075102
\bibitem{Solyom} S\'{o}lyom J (1979) {\it Adv.~Phys.} {\bf 28} 201
\bibitem{DMRGrev} For a recent review see: 
Schollw\"ock U (2005) {\it Rev.~Mod.~Phys.} {\bf 77} 259  
\bibitem{Claudio} Chamon C, Oshikawa M, and Affleck I (2003)
  {\it Phys.\ Rev.\ Lett.} {\bf 91} 206403
\bibitem{Aristov} Aristov D N and W\"olfle P (2007) arXiv:0709.1562
\bibitem{Taylor} Taylor J R (1972) {\it Scattering Theory}, John Wiley and
  Sons, New York

\end{thebibliography}
\end{document}